\documentclass[fleqn,usenatbib]{mnras}

\usepackage[T1]{fontenc}
\usepackage{ae,aecompl}

\usepackage{graphicx}	
\usepackage{amsmath}	
\usepackage{amssymb}	
\usepackage{times}

\newcommand{\bs}{\boldsymbol}
\newcommand{\percent}{\,\mathrm{per\,cent}}
\usepackage{array}
\newcolumntype{x}[1]{>{\centering\arraybackslash\hspace{0pt}}p{#1}}
\usepackage{newtxtext,newtxmath}

\title[Extinction Law in the Inner Galaxy]{The extinction law in the inner $3\times3\deg^2$ of the Milky Way and the red clump absolute magnitude in the inner bar-bulge}

\author[J. L. Sanders et al.]{Jason L. Sanders,$^{1}$\thanks{E-mail: jason.sanders@ucl.ac.uk (JLS)}
Leigh C. Smith,$^{2}$
Carlos Gonz\'alez-Fern\'andez,$^{2}$
Philip Lucas,$^{3}$
\newauthor
Dante Minniti$^{4,5}$
\\
$^{1}$Department of Physics and Astronomy, University College London, London WC1E 6BT, UK\\
$^{2}$Institute of Astronomy, University of Cambridge, Madingley Rise, Cambridge, CB3 0HA, UK\\
$^{3}$Centre for Astrophysics, University of Hertfordshire, College Lane, Hatfield AL10 9AB, UK\\
$^{4}$Departamento de Ciencias F\'isicas, Facultad de Ciencias Exactas, Universidad Andr\'es Bello, Fern\'andez Concha 700, Las Condes, Santiago, Chile\\
$^{5}$Vatican Observatory, Vatican City State, V-00120, Italy
}

\date{Accepted XXX. Received YYY; in original form ZZZ}

\pubyear{2021}

\begin{document}
\label{firstpage}
\pagerange{\pageref{firstpage}--\pageref{lastpage}}
\maketitle

\begin{abstract}
The extinction law from $0.9$ to $8$ microns in the inner $3\times3\deg^2$ of the Milky Way is measured using data from VISTA Variables in the Via Lactea, GLIMPSE and WISE. Absolute extinction ratios are found by requiring that the observed red clump density peaks at the GRAVITY collaboration distance to the Galactic centre. When combined with selective extinction ratios measured from the bulge giant colour-colour diagrams, we find an extinction law of $A_Z:A_Y:A_J:A_H:A_{K_s}:A_{W1}:A_{[3.6]}:A_{[4.5]}:A_{W2}:A_{[5.8]}:A_{[8.0]} =7.19(0.30):5.11(0.20):3.23(0.11):1.77(0.04):1:0.54(0.02):0.46(0.03):0.34(0.03):0.32(0.03):0.24(0.04):0.28(0.03)$ valid for low extinctions where non-linearities are unimportant. These results imply an extinction law from the Rayleigh Jeans colour excess method (RJCE) of $A_{K_s}=0.677(H-[4.5]-0.188)$. We find little evidence for significant selective extinction ratio variation over the inspected region (around $5\percent$). Assuming the absolute extinction ratios do not vary across the inspected region gives an independent measurement of the absolute $K_s$ magnitude of the red clump at the Galactic Centre of $(-1.61\pm0.07)\,\mathrm{mag}$. This is very similar to the value measured for solar neighbourhood red clump stars giving confidence in the use of red clump stars as standard candles across the Galaxy. As part of our analysis, we inspect the completeness of PSF photometry from the VVV survey using artificial star tests, finding $90\percent$ completeness at $K_s\approx16 \,(17)$ in high (low) density regions and good agreement with the number counts with respect to the GALACTICNUCLEUS and DECAPS catalogues over small regions of the survey.
\end{abstract}



\section{Introduction}
The inner parts of the Milky Way contain important information on the formation and history of our Galaxy and provide a detailed window into the nuclear properties of a Milky-Way-like galaxy. Near- and mid-infrared surveys such as 2MASS, VVV, UKIDSS and GLIMPSE have allowed studies to probe through the interstellar dust to infer the morphological properties of this region. On scales of $\ell\lesssim10\,\mathrm{deg}$ the Galactic bar/bulge is the dominant structure \citep[e.g.][]{BlitzSpergel1991,Stanek1994,Stanek1997,BabusiauxGilmore2005,Rattenbury2007,McWilliamZoccali2010,WeggGerhard2013,Simion2017,Sanders2019a}, and at larger radii, the bar/bulge extends into the long thin bar \citep{Wegg2015}. In the very central regions around the suspected dynamical centre of the Galaxy, Sgr A*, there is a nuclear stellar cluster \citep[NSC,][]{Schoedel2014}, and between these two structures sits the nuclear stellar disc \citep{Launhardt2002,Nishiyama2013,Sormani2021}. The nuclear stellar disc (NSD) has an extent of $\ell\sim1.5\,\mathrm{deg}$ and is associated with the central x2 orbits that form in a barred potential. Compared to the bar/bulge and the NSC, the NSD has been little studied. \cite{NoguerasLara2020} recently argued that the nuclear stellar disc is composed almost entirely of stars older than $8\,\mathrm{Gyr}$ although there is evidence of a more recent starburst. This points towards a picture where the bulk of the stars in the nuclear stellar disc formed at bar formation when gas rapidly funnelled to the centre, but since then the gas supply and hence star formation has been more gentle until possibly a recent episode.

Many of the highlighted studies on the structure of the inner Galaxy rely on the standard candle nature of red clump stars \citep{GirardiReview}. Red clump stars are metal-rich core helium burning and, particularly in the $K_s$ band, display a small scatter in absolute magnitude \citep{Alves2000,Hawkins2017,Hall2019,ChanBovy2020}. For this reason, they have found use beyond the Milky Way, for example in the Magellanic Clouds \citep{Alves2002,Pietrzynski2002} and M31 \citep{Stanek1998}. Variations in the red clump's absolute magnitude with both age and metallicity are expected \citep{GirardiSalaris2001} and observed \citep{Chen2017,Huang2020,ChanBovy2020}. This means for highly accurate density mapping of the central Galaxy we require knowledge of its age and metallicity structure. Alternatively, with a known density structure we could use the red clump to infer the age and metallicity of the inner Galaxy \citep{NoguerasLara2020}.

These approaches are complicated by extinction, which is particularly severe in the inner regions of the Milky Way \citep[$A_V\approx40$ towards the Galactic Centre,][]{Nishiyama2008,Schodel2010}. In the absence of additional information, there are degeneracies between the absolute magnitude of the red clump, the distances to the stars and the shape of the extinction curve. For structural studies of the inner Galaxy, the extinction curve shape is a key parameter.  \cite{Matsunaga2018} gives a review of measurements of the extinction law in the near- and mid-infrared. In recent years, studies have settled on a steep extinction law in the near-infrared with power-law index $\alpha$ ($A_x\propto\lambda_x^{-\alpha}$) around $\alpha=2-2.2$ \citep{SteadHoare2009,Nishiyama2009,Schodel2010,Fritz2011,AlonsoGarcia2017,NoguerasLara2019,MaizApellaniz2020,StelterEikenberry2021}. The universality of the extinction law is less certain with variations reported as a function of location in the Galaxy \citep{Zasowski2009} and total extinction/density of intervening clouds \citep{Chapman2009}, although \cite{WangJiang2014} argue for a universality of the near-infrared extinction law.

Until the measurements of the distance to Sgr A* from \cite{GravityCollaboration2019,GravityCollaboration2021}, a reference distance to the bar/bulge red clump stars was not known. Instead several studies have inferred the distance to the Galactic Centre using assumptions for the red clump absolute magnitude based on observations of solar neighbourhood stars \citep{PaczynskiStanek,Nishiyama2006}. The degeneracies between absolute magnitude, distance and extinction make these studies challenging. However, with the distance to the Galactic Centre now more confidently measured, we are in a position to break some of this degeneracy under the assumption that Sgr A* is the dynamical centre of the Galaxy.

In this paper, we derive the extinction curve of the inner $3\times3\,\mathrm{deg}^2$ of the Galaxy from $0.9$ to $8$ microns. We use a novel method that pins the unextincted magnitude of the red clump based on the known distance to the Galactic Centre from \cite{GravityCollaboration2021}. In this way, we are able to not only infer the extinction law but also the absolute magnitude of red clump stars at the Galactic centre. We combine this method with more traditional colour excess methods to derive the full extinction curve. We outline and apply the colour excess method in Section~\ref{section::extinction}, and the calibration of the absolute extinction curve and a discussion of the absolute red clump magnitude are presented in Section~\ref{section::abs}. From this extinction curve, we derive $2$D extinction maps (neglecting distance dependence) using both the known colours of red clump stars \citep{Gonzalez2011,Gonzalez2012,Surot2020} and the $(H-[4.5])$ colour of all stars \citep{Majewski2011} as presented in Appendix~\ref{appendix::rc_method} and~\ref{appendix::rjce_method}. We further present a brief discussion of the completeness properties of a point-spread-function photometric reduction of VVV in Appendix~\ref{section::completeness}. We close with our conclusions in Section~\ref{sec::conclusions}.

\section{Selective extinction ratios from 1 to 8 microns}\label{section::extinction}
In this section, we consider measuring the selective extinction ratios for the inner Galaxy. It is much easier to estimate selective extinction ratios than any absolute to selective extinction ratio as we only require observed colours of known stellar types, not knowledge of the distances to the sources. However, with a set of selective ratios and a single absolute to selective ratio, we can compute all absolute extinction ratios. 

An extinction law $A(\lambda)$ describes the reduction in magnitudes at a given wavelength $\lambda$. The extinction in a photometric band $x$ with (photon-counting) transmission $T_x(\lambda)$ for a star with flux $F_\lambda(\lambda)$ is
\begin{equation}
    A_x = -2.5\log_{10}\Big(\int\mathrm{d}\lambda\,\lambda T_x F_\lambda 10^{-0.4A(\lambda)}\Big/\int\mathrm{d}\lambda\,\lambda T_xF_\lambda\Big),
\label{eqn::ext_integral}
\end{equation}
from which we can define the selective extinction for colour $x-y$ as $E(x-y)\equiv A_x-A_y$. For low extinction, the integral is approximately linear in $A(\lambda)$ and for narrow bands $A_x$ is approximately independent of $F_\lambda(\lambda)$. In these limits, the extinction coefficients normalized with respect to the extinction at some reference wavelength, $A_x/A(\lambda_\mathrm{ref})$, are independent of total extinction and stellar type. In this case, a universal extinction law can be applied to all stars and we characterise the extinction law by constant absolute ratios e.g. $A_x/A_{K_s}$ from which constant selective extinctions ratios can be computed e.g. $E(x-K_s)/E(y-K_s)=(A_x/A_{K_s}-1)/(A_y/A_{K_s}-1)$.

Assuming non-linear effects are minimal, for some colour $c$, the extinction law is given by
\begin{equation}
    A_{K_s} = R_c E(c),
\end{equation}
for a constant $R_c$. Our aim is to measure $E(c)/E(H-K_s)=R_{H-K_s}/R_c$.
We consider red giant stars which are known to follow trends in colour-magnitude space related to correlations between luminosity and temperature. As our samples are located predominantly in the Galactic bulge, these correlations are imprinted on the dereddened colour magnitude diagrams. It is therefore important to consider this `intrinsic' correlation in addition to correlations induced by extinction. Assuming the absolute $K_s$ magnitude is a simple linear function of colour for giant stars at a single distance, we have
\begin{equation}
    K_s' = (c'-c_0)/k_c+K_{s0,c},
\end{equation}
where primed quantities are dereddened and $c_0$ and $K_{s0,c}$ are arbitrary constants. We measure $k_c$ from a set of PARSEC isochrones with $\mathrm{[M/H]}=0$ and $\log(\mathrm{age/Gyr})>9.5$ \citep{Bressan2012,Chen2014a,Chen2015,Tang2014,Marigo2017}. Using these relations we can relate the observed $(H-K_s)$ colour to any other colour, $c$, as
\begin{equation}
    (H-K_s) - k_{H-K_s}K_s = \frac{R_c}{R_{H-K_s}}\frac{1-k_{H-K_s}R_{H-K_s}}{1-k_cR_c}(c-k_c K_s)+\mathrm{const.},
\label{eqn::selective}
\end{equation}
so the gradient $g_c$ of a linear fit of $c-k_cK_s$ against $(H-K_s)-k_{H-K_s}K_s$  will recover the extinction ratio $E(c)/E(H-K_s)$ as
\begin{equation}
    \frac{E(c)}{E(H-K_s)} = g_c+R_{H-K_s}(k_c-g_ck_{H-K_s}).
\end{equation}
Note that this method requires knowledge of the absolute extinction ratio $R_{H-K_s}$ to measure $E(c)/E(H-K_s)$. However, the dependence is weak as $k_c$ for near- and mid-infrared bands is typically small so we adopt a standard value from \cite{Fritz2011} of $R_{H-K_s}=1.328$ (using $R_{H-K_s}=1.104$ from \cite{AlonsoGarcia2017} results in changes to $E(c)/E(H-K_s)$ of less than $0.5\percent$). We note that it is possible to use a version of equation~\eqref{eqn::selective} dependent on $E(H-K_s)$ measured from red clump stars. However, we find in general this method is less accurate than using colour-colour relations, partly because of the underestimate of $E(H-K_s)_\mathrm{RC}$ for large extinction (see Appendix~\ref{appendix::rc_method}). 

To fit equation~\eqref{eqn::selective} to data we employ a total least-squares procedure \citep{Hogg2010}. For a data vector $\bs{Z}_i = (c_i-k_c K_{si}, (H-K_s)_i - k_{H-K_s}K_{si})$ and corresponding covariance matrix $\bs{\Sigma}_i$ we seek the straight line that maximises the log-likelihood
\begin{equation}
    \ln\mathcal{L} = -\tfrac{1}{2}\sum_i \ln(\bs{v}^{\rm T}\Sigma_i\bs{v}+\sigma^2)+\frac{(\bs{v}^{\rm T} \bs{Z}_i-b_\perp)^2}{\bs{v}^{\rm T}\Sigma_i\bs{v}+\sigma^2},
\end{equation}
where $\bs{v}=(-\sin\theta,\cos\theta)$, $\theta$ is the angle of the line with respect to the `$x$' coordinate, $b_\perp=b\cos\theta$ with $b$ the `$y$' intercept of the line. We adopt flat priors in $\theta$ and $b_\perp$. We add in an additional variance $\sigma^2$ to account for the width of the giant branch. Additionally, we choose to include an outlier component which a different straight line with a broader variance $\sigma_\mathrm{out}^2$. The weight of the outlier component is $f_\mathrm{out}$ which we limit to be less than $0.4$. We sum the likelihood of this term with the model likelihood (times $(1-f_\mathrm{out})$). The maximum likelihood is found through using BFGS minimization with the Jacobian computed using \textsc{autograd} \footnote{\url{https://github.com/HIPS/autograd}}. The uncertainty in the output gradient is estimated using the inverse of the Hessian matrix (also computed using \textsc{autograd}).

\subsection{Non-linearity of extinction coefficients}\label{section::nonlinearity}
\begin{figure}
    \includegraphics[width=\columnwidth]{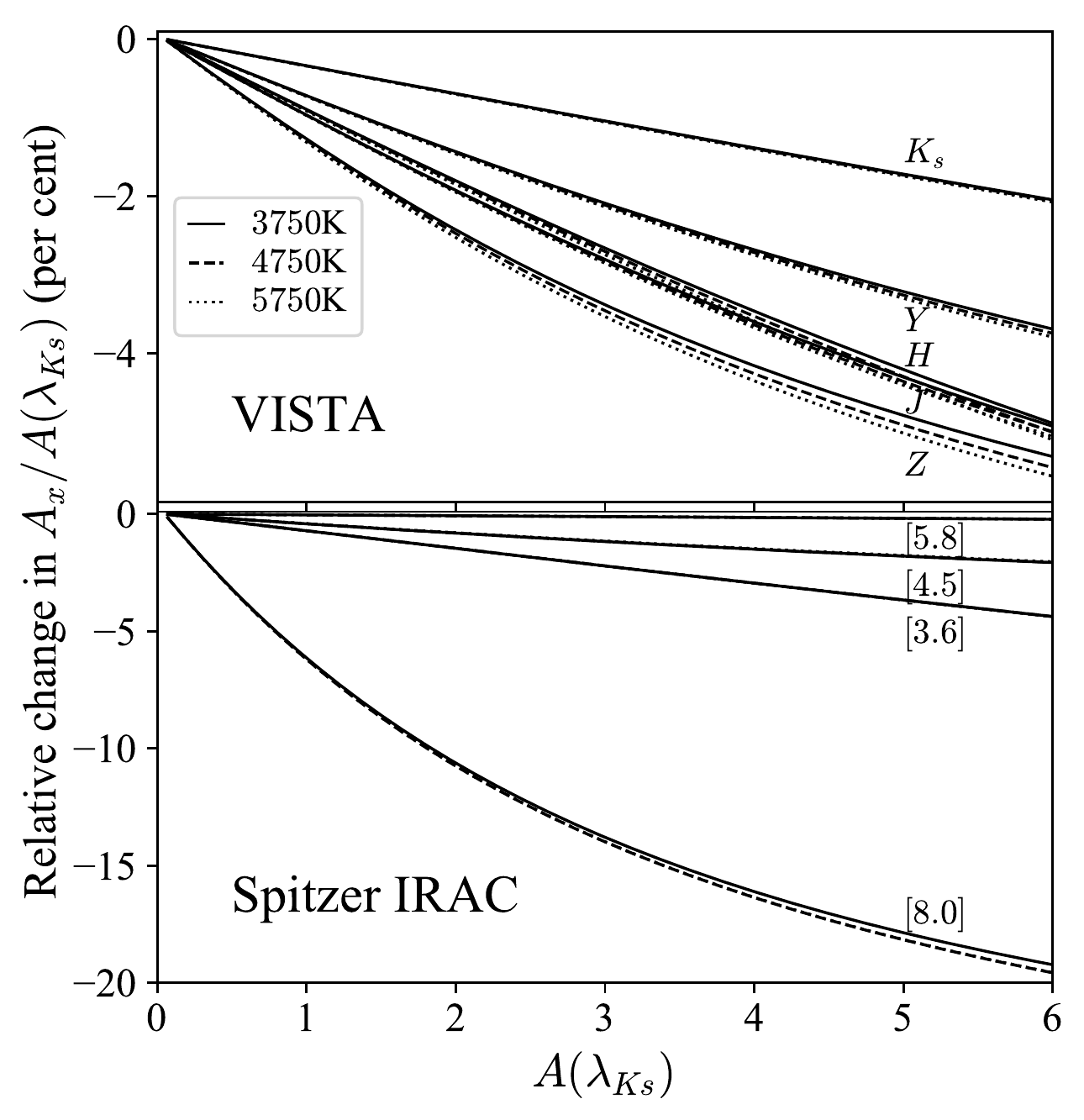}
    \caption{Non-linearity of the extinction law: we display curves of the relative percentage change in the extinction in band $x$, $A_x$ (evaluated using equation~\eqref{eqn::ext_integral}), normalized by the extinction evaluated at $\lambda_{K_s}=2.149\,\mu\mathrm{m}$. We adopt the extinction curve from \protect\cite{Fritz2011} and use three different giant stellar spectra.}
    \label{fig::nonlinear}
\end{figure}
For high extinction regions, the non-linearity in the extinction calculation of equation~\eqref{eqn::ext_integral} can become important \citep[e.g.][]{SteadHoare2009,WangChen2019,MaizApellaniz2020}. We assess the impact of the non-linearity by computing equation~\eqref{eqn::ext_integral} for the VISTA and Spitzer passbands \citep[downloaded from the SVO filter service,][]{svo1,svo2}. We use three different stellar spectra from \cite{CastelliKurucz2004} with $T_\mathrm{eff}=(3750,4750,5750)\,\mathrm{K}$, $\log g=(0.5,2.5,3.5)$ and solar metallicity. We use the extinction law from \cite{Fritz2011}. In Fig.~\ref{fig::nonlinear} we display the percentage relative change in $A_x$ normalized by the extinction at $\lambda_{K_s}=2.149\,\mu\mathrm{m}$, $A_{K_s}$, against $A_{K_s}$. Most bands have relative changes of a few per cent over the range of considered extinctions ($A(\lambda_{K_s})<6$, typically extinctions in the inner Galaxy are $A(\lambda_{K_s}<3$) and in general, the bluer VISTA bands have larger changes than the Spitzer bands, except for
$[8.0]$ for which there is a large evolution of $A_{[8.0]}/A(\lambda_{K_s})$ of approximately $5\percent$ per magnitude of $A(\lambda_{K_s})$. This arises due to the $9.7\,\mu\mathrm{m}$ silicate feature. From these calculations, we store the variation of $\mathrm{NL}_c(E(H-K_s)) = E(c)/E(H-K_s) - E(c)/E(H-K_s)|_{A_x\rightarrow0}$ as a function of $E(H-K_s)$ for the $T_\mathrm{eff}=4750\,\mathrm{K},\,\log g=2.5$ model to simply correct the colours of stars for typical non-linear effects as $c \leftarrow c-E(H-K_s)\mathrm{NL}_c(E(H-K_s))$ using $E(H-K_s)\approx(H-K_s-0.1)$.

\subsection{Data}\label{sec::data}
\begin{figure}
    \centering
    \includegraphics[width=\columnwidth]{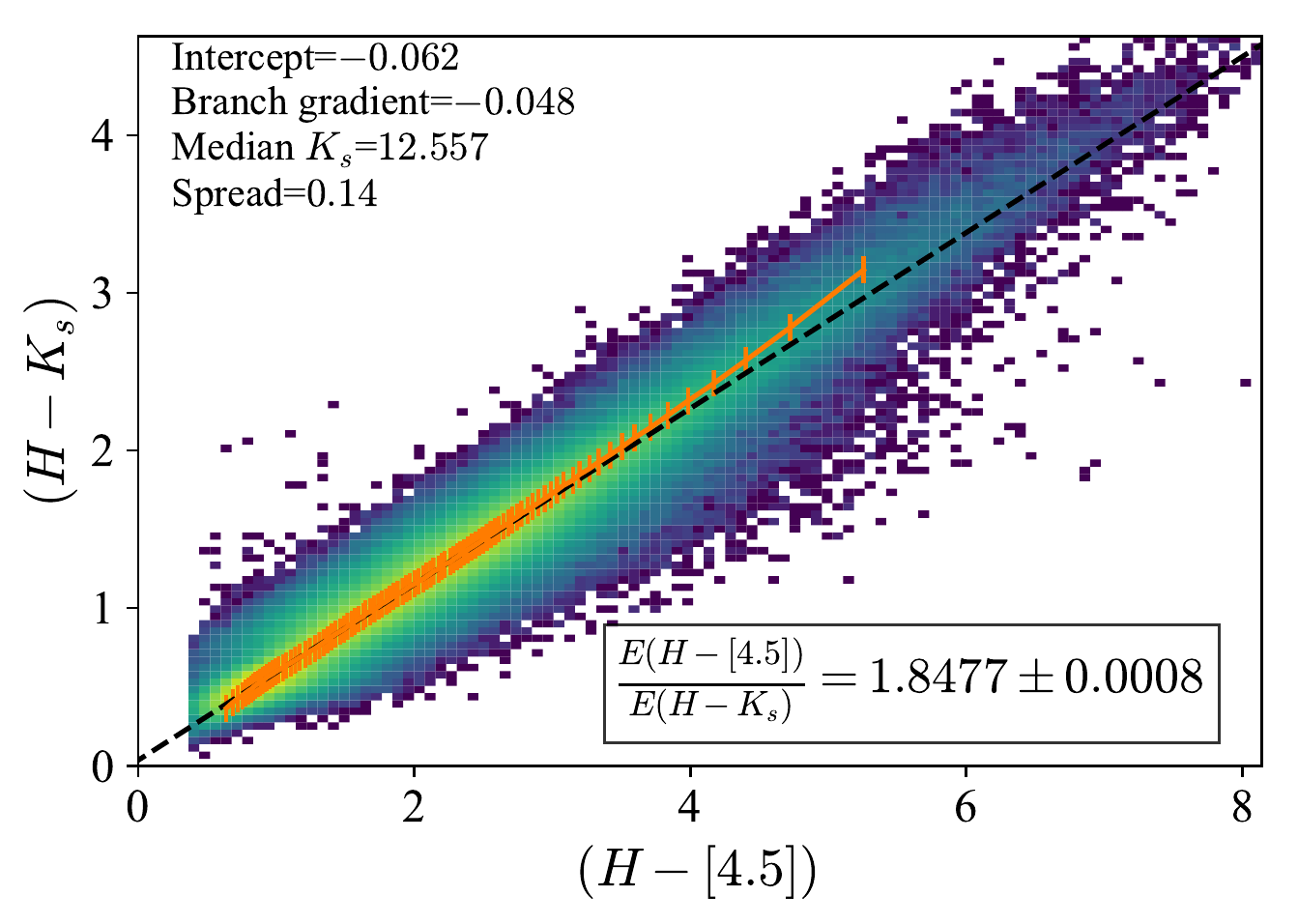}
    \caption{Example of a colour-colour giant branch fit to find the selective extinction ratio for $E(H-[4.5])/E(H-K_s)$. The background histogram shows the full data distribution coloured using a log-scale. The orange is the binned median, whilst the black-dashed shows a linear fit using a total least-squares approach. The `intercept' gives the x-intercept of the fit, `branch gradient' is the slope of the giant branch in $(H-[4.5]$ vs. $K_s$, and `spread' is the standard deviation about the linear fit.}
    \label{fig::example}
\end{figure}
Our primary source of data is the preliminary VIRAC2 catalogue \citep[][Smith et al. in prep.]{VIRAC}. VIRAC2 is an updated version of the VVV InfraRed Astrometric Catalogue presented by \cite{VIRAC}.
VVV is a near-infrared multi-epoch photometric survey of the Galactic bulge and southern Galactic disc performed using the VIRCAM detector \citep{Dalton2006} with pixel resolution of $0.339\,\mathrm{arcsec}$ on the $4\,\mathrm{m}$ VISTA telescope at the Cerro Paranal Observatory (with typical seeing of $\sim0.7\,\mathrm{arcsec}$). The original VVV survey ran from $2010$ to $2015$ but was granted a five-year extension (VVVX) from $2016$. The primary multi-epoch observations consist of $\sim200$ $2\times4\,\mathrm{s}$ $K_s$ exposures with a typical depth of $K_s\sim17$ in the region we consider (each pointing consists of two exposures `jittered' by small offsets, hence the `$2\times$'). The $K_s$ photometry is complemented by more sparsely sampled $Z$, $Y$, $J$ and $H$ photometry (approximately $4$, $4$, $20$--$40$ and $10$--$20$ exposures of $2\times10$, $2\times10$, $4\times6$ ($4\times10$ for VVVX) and $2\times4$ ($2\times6$ for VVVX) seconds respectively, see \citealt{Saito2012} for details). There are no $Z$ and $Y$ observations as part of VVVX. Saturation effects in all bands occur around $11.5$ to $13.0\,\mathrm{mag}$ \citep[see figure 12 of][for results with $5$s exposures]{GonzalezFernandez2018} and the typical depth is $(20.5,20,19.5,18.5,18)\,\mathrm{mag}$ in $(Z,Y,J,H,K_s)$ (slightly deeper for the longer VVVX $J$ and $H$ exposures) although in the central bulge fields the depth is confusion-limited \citep[see figures 3, 4 and 5 of][]{Saito2012}.

VIRAC2 consists of a complete re-reduction of the VVV and VVVX $ZYJHK_s$ images in the original VVV Galactic bulge and southern disc footprint utilising point-spread-function photometry, employing a recalibration of the photometric zeropoints with respect to 2MASS \citep[to address the issues highlighted by][]{Hajdu2020} and combining multiple detections into astrometric solutions calibrated with respect to Gaia. Furthermore, as VVV begins to saturate around $K_s=11.5$, we complement the VVV dataset with brighter stars from 2MASS by finding any star in 2MASS without a counterpart in VIRAC2 within $1\,\mathrm{arcsec}$ and with $K_s>11$. We put the 2MASS measurements on the VISTA system using the transformations from \cite{GonzalezFernandez2018} and correct for small median photometric shifts in overlapping 2MASS and VIRAC2 measurements. We perform our procedure described in the previous section on all stars in VIRAC2 (combined with 2MASS) with $|\ell|<1.5\,\mathrm{deg}$, $|b|<1.5\,\mathrm{deg}$ and $K_s<13.5-1.33(H-K_s-0.09)$ (a cut approximately parallel to the extinction vector), five-parameter astrometric solutions, non-duplicates and without a neighbour in the dataset within $1\,\mathrm{arcsec}$. When using VIRAC2 $ZYJHK_s$ photometry we limit to magnitude uncertainties $<0.06\,\mathrm{mag}$ and magnitudes brighter than $19$. We cross-match the sources to sources in GLIMPSE (using sources with \texttt{sqf\_*}$=0$ from \citealt{GLIMPSE} or \texttt{f*}$=1$ from \citealt{Ramirez2008}) and unWISE \citep[][using sources with no flags]{unWISE} with a cross-match radius of $0.4\,\mathrm{arcsec}$ and $1\,\mathrm{arcsec}$ respectively (again limiting to stars with magnitude uncertainties $<0.2\,\mathrm{mag}$). Our other cuts, in particular the VIRAC2 nearest neighbour cut, are such that essentially all the VIRAC2 sources we use are assigned to unique (or no) WISE/GLIMPSE sources. One source is cross-matched to two unWISE sources. When using WISE, we also ensure there is a corresponding GLIMPSE detection. To eliminate foreground contaminants, we further remove stars with $(H-K_s)+2\sqrt{\sigma_H^2+\sigma_{Ks}^2}<0.09+E(H-K_s)_\mathrm{RC}-2\sigma_{E(H-K_s)_\mathrm{RC}}$ or $(J-K_s)+2\sqrt{\sigma_J^2+\sigma_{Ks}^2}<0.62+E(J-K_s)_\mathrm{RC}-2\sigma_{E(J-K_s)_\mathrm{RC}}$ using extinction estimates from the red clump method (Appendix~\ref{appendix::rc_method}). Note this procedure is valid even when $E(J-K_s)_\mathrm{RC}$ and $E(H-K_s)_\mathrm{RC}$ are biased low due to the absence of the red clump in high extinction regions. We further remove likely AGB or YSOs by removing stars with $([5.8]-[8.0])/\sqrt{\sigma_{[5.8]}^2+\sigma_{[8.0]}^2}>5$. The extinction coefficient in $[5.8]$ is smaller than $[8.0]$ \citep[see][]{WangChen2019} so this does not remove highly extincted giant stars. Cross-matching this sample to GALACTICNUCLEUS \citep[][with a $0.1\,\mathrm{arcsec}$ cross-match radius]{GALACTICNUCLEUS} and restricting to stars with $K_s<14$ and VIRAC uncertainties $<0.02\,\mathrm{mag}$ gives zero-point offsets (GALACTICNUCLEUS - VIRAC) of $(J,H,K_s)=(-0.027,0.008,0.026)$.

For $x=\{Z, Y, J, [3.6], [4.5], [5.8], [8.0], W1, W2\}$ with $c=(x-H)$, we find the maximum likelihood solution for the straight line fit as described earlier.
The result of applying our procedure with $x=[4.5]$ is shown in Fig.~\ref{fig::example} where we measure $E(H-[4.5])/E(H-K_s)=1.8477\pm0.0008$ and $k_{[4.5]-H}=-0.021$. 
We present the results of our fitting procedure in Table~\ref{tab:colour_ratio_results}. Note that the uncertainties are statistical and probably unrealistically small as they do not reflect systematic uncertainty from the simplicity of the employed model. We compare to the ratios to reported by \citet[][using hydrogen line absorption towards the Galactic Centre]{Fritz2011}, \citet[][using the colour-magnitude slope of the bulge red clump stars]{AlonsoGarcia2017}, \citet[][using the colour-magnitude slope of giant stars in the bulge]{Nishiyama2009} and \citet[][using red clump stars from APOGEE distributed across the sky]{WangChen2019}. It should be noted that as different studies use different instruments the infra-red bands are not identical so some variation is to be expected. Using the $T_\mathrm{eff}=4750\,\mathrm{K}, \log g=2.5$ spectrum from \cite{CastelliKurucz2004}, we find $(A_{J_\mathrm{HAWK}},A_{J_\mathrm{Sirius}},A_{J_\mathrm{2MASS}})=(0.994, 1.012, 1.028)A_{J_\mathrm{VISTA}}$, $(A_{H_\mathrm{HAWK}},A_{H_\mathrm{Sirius}},A_{H_\mathrm{2MASS}})=(1.037, 1.021, 0.995)A_{H_\mathrm{VISTA}}$ and $(A_{K_{s,\mathrm{HAWK}}},A_{K_{s,\mathrm{Sirius}}},A_{K_{s,\mathrm{2MASS}}})=(1.012, 1.011, 0.991)A_{K_{s,\mathrm{VISTA}}}$.

\begin{table*}
    \caption{Colour excess ratio $E(x-H)/E(H-K_s)$ (left for the full region, right split into subpixels with the uncertainty showing the standard deviation) with the corresponding measurements from \protect\cite{Fritz2011}, \protect\cite{AlonsoGarcia2017}, \protect\cite{Nishiyama2009}, \protect\cite{NoguerasLara2019} and \protect\cite{WangChen2019}. \protect\cite{WangChen2019} use the Spitzer ratios from \protect\cite{Chen2018}.  Note that \protect\cite{Nishiyama2009} report coefficients for the SIRIUS bands,
    \protect\cite{WangChen2019} for the 2MASS bands
    and \protect\cite{NoguerasLara2019} for the HAWK bands (see the end of Section~\ref{sec::data} for approximate conversions).
    }
    \centering
    \setlength\tabcolsep{3pt}
    \begin{tabular}{lccx{0.22\columnwidth}x{0.22\columnwidth}x{0.2\columnwidth}x{0.22\columnwidth}x{0.22\columnwidth}}
    \hline
    $x-H$&\multicolumn{2}{c}{$E(x-H)/E(H-K_s)$}&\cite{Fritz2011}&\cite{AlonsoGarcia2017}&\cite{Nishiyama2009}&\cite{NoguerasLara2019}&\cite{WangChen2019}\\
    \hline
$Z$&$+7.0010\pm0.0018$&$+7.07\pm0.25$&$-$&$+6.66\pm0.23$&$-$&$-$&$-$\\
$Y$&$+4.3119\pm0.0007$&$+4.38\pm0.18$&$+3.84\pm0.43$&$+3.98\pm0.13$&$-$&$-$&$-$\\
$J$&$+1.8788\pm0.0001$&$+1.98\pm0.08$&$+1.76\pm0.20$&$+1.61\pm0.05$&$+1.76\pm0.06$&$+2.09\pm0.05$&$+2.11\pm0.31$\\
$[3.6]$&$-1.7034\pm0.0008$&$-1.70\pm0.04$&$-1.60\pm0.30$&$-$&$-1.68\pm0.11$&$-$&$-1.77\pm0.47$\\
$[4.5]$&$-1.8477\pm0.0008$&$-1.82\pm0.07$&$-1.80\pm0.33$&$-$&$-1.83\pm0.12$&$-$&$-1.98\pm0.51$\\
$[5.8]$&$-1.9861\pm0.0006$&$-1.97\pm0.06$&$-1.88\pm0.34$&$-$&$-1.88\pm0.12$&$-$&$-2.11\pm0.53$\\
$[8.0]$&$-1.9264\pm0.0007$&$-1.87\pm0.08$&$-1.82\pm0.35$&$-$&$-1.78\pm0.11$&$-$&$-2.00\pm0.51$\\
$W1$&$-1.5892\pm0.0044$&$-1.57\pm0.05$&$-$&$-$&$-$&$-$&$-1.74\pm0.47$\\
$W2$&$-1.8789\pm0.0053$&$-1.86\pm0.05$&$-$&$-$&$-$&$-$&$-1.98\pm0.51$\\
    \hline
    \end{tabular}
    \label{tab:colour_ratio_results}
\end{table*}

\subsection{Spatial variation}
\begin{figure*}
    \centering
    \includegraphics[width=\textwidth]{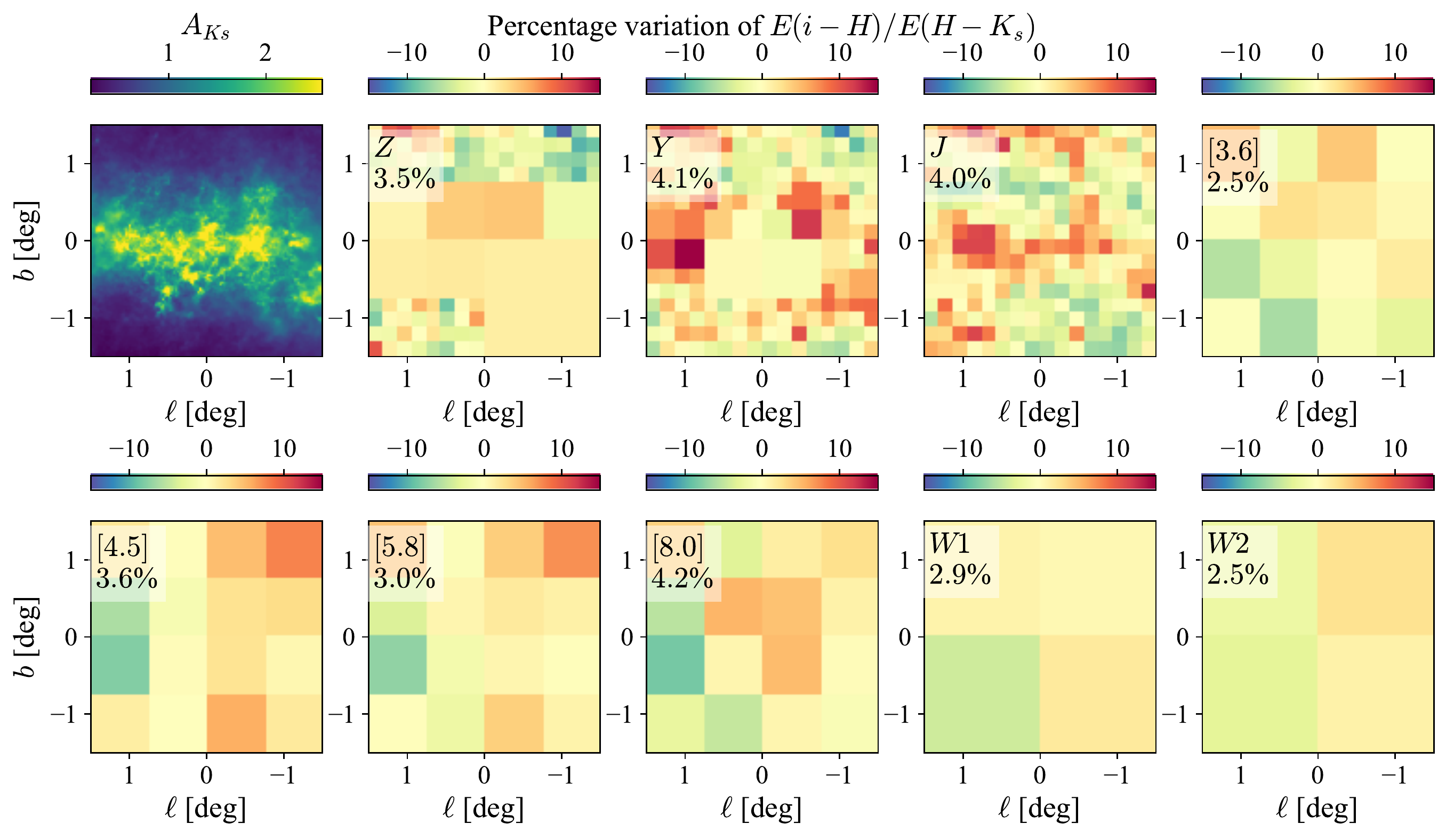}
    \caption{Percentage variation in the selective extinction ratios $E(i-H)/E(H-K_s)$. Top left shows the $A_{K_s}$ extinction map computed using the RJCE method (see Appendix~\ref{appendix::rjce_method}). Pixels are subdivided to ensure at least $500$ stars per pixel. 
    Typically variations are $\lesssim5\percent$.}
    \label{fig:spatial_variation}
\end{figure*}
With the average selective extinction ratios over the inner $3\times3\,\mathrm{deg}^2$ computed, we now turn to the spatial variation of the selective extinction ratio. We repeat the above calculation for a series of on-sky bins. Starting with the whole considered region, we iteratively split the on-sky bins into four sub-bins until further splits would cause one sub-bin to have fewer than $500$ stars or a total of four splits have already been performed (or two splits when considering Spitzer/GLIMPSE data). We adopt a stricter cut on Spitzer/GLIMPSE photometry uncertainties requiring magnitude errors smaller than $0.05\,\mathrm{mag}$.

We display the results in Fig.~\ref{fig:spatial_variation} and in Table~\ref{tab:colour_ratio_results} we report the mean and standard deviation of the selective extinction ratios across the on-sky bins. In all bands, we find less than $5$ percent  variation in $E(i-K_s)/E(H-K_s)$ across the region analysed, which translates into a few percent variation in $A_i/A_{K_s}$ for the bluer bands ($Z$, $Y$, $J$) and a few tens of percent for the redder bands (e.g. $[8.0]$). Some part of the variation appears as noise, but there are also some clear systematic variations. For instance, there is a feature of the presented maps around $(\ell,b)=(1,0)$ seen particularly in $J$ and $Y$, but also seen in the Spitzer bands as a general underestimate. This feature also coincides with a problematic region for the 2D extinction maps (see Appendix~\ref{appendix::rjce_method} and Fig.~\ref{fig::rjce_comparison}) where it appears there is a significant population of young stars in this region which bias low the extinction to the red clump stars. This suggests along these lines-of-sight there is significant extinction variation. Foreground clouds could give rise to variation in the selective extinction ratios \citep[e.g.][]{Chapman2009}. However, the variation we see could also be attributed to the contaminating young star population.  \cite{Nishiyama2009}, \cite{AlonsoGarcia2017} and \cite{NoguerasLara2019} all investigated the variation of the extinction law across the Galactic central regions. \cite{AlonsoGarcia2017} tabulated the variation in the extinction law for four quadrants in the region $|l|\lesssim2.7\,\mathrm{deg}$ and $|b|\lesssim1.55\,\mathrm{deg}$ finding variation of order $\sim10\percent$ for $E(i-H)/E(H-K_s)$ for $i\in\{Z,Y,J\}$ and in general higher values for the southern quadrants compared to northern. \cite{Nishiyama2009} found a similar $10\percent$ variation in $E(J-H)/E(H-K_s)$ across the region $|l|\lesssim3\,\mathrm{deg}$ and $|b|\lesssim1\,\mathrm{deg}$ but instead with higher values observed in the northern quadrants (although as \citealt{AlonsoGarcia2017} report, the trend of lower $A_{K_s}/E(J-K_s)$ in the northern quadrants was found in both studies).
\cite{NoguerasLara2019} inspected the variation across the nuclear stellar disc ($|l|\lesssim0.3\,\mathrm{deg}$ and $|b|\lesssim0.1\,\mathrm{deg}$) finding no significant variation in the slope of the extinction law. In conclusion, we find that the variation in the selective extinction law across the inner $3\times3\,\mathrm{deg}^2$ is small.

\section{Absolute extinction ratios and the magnitude of the red clump}\label{section::abs}
\begin{figure}
    \centering
    \includegraphics[width=\columnwidth]{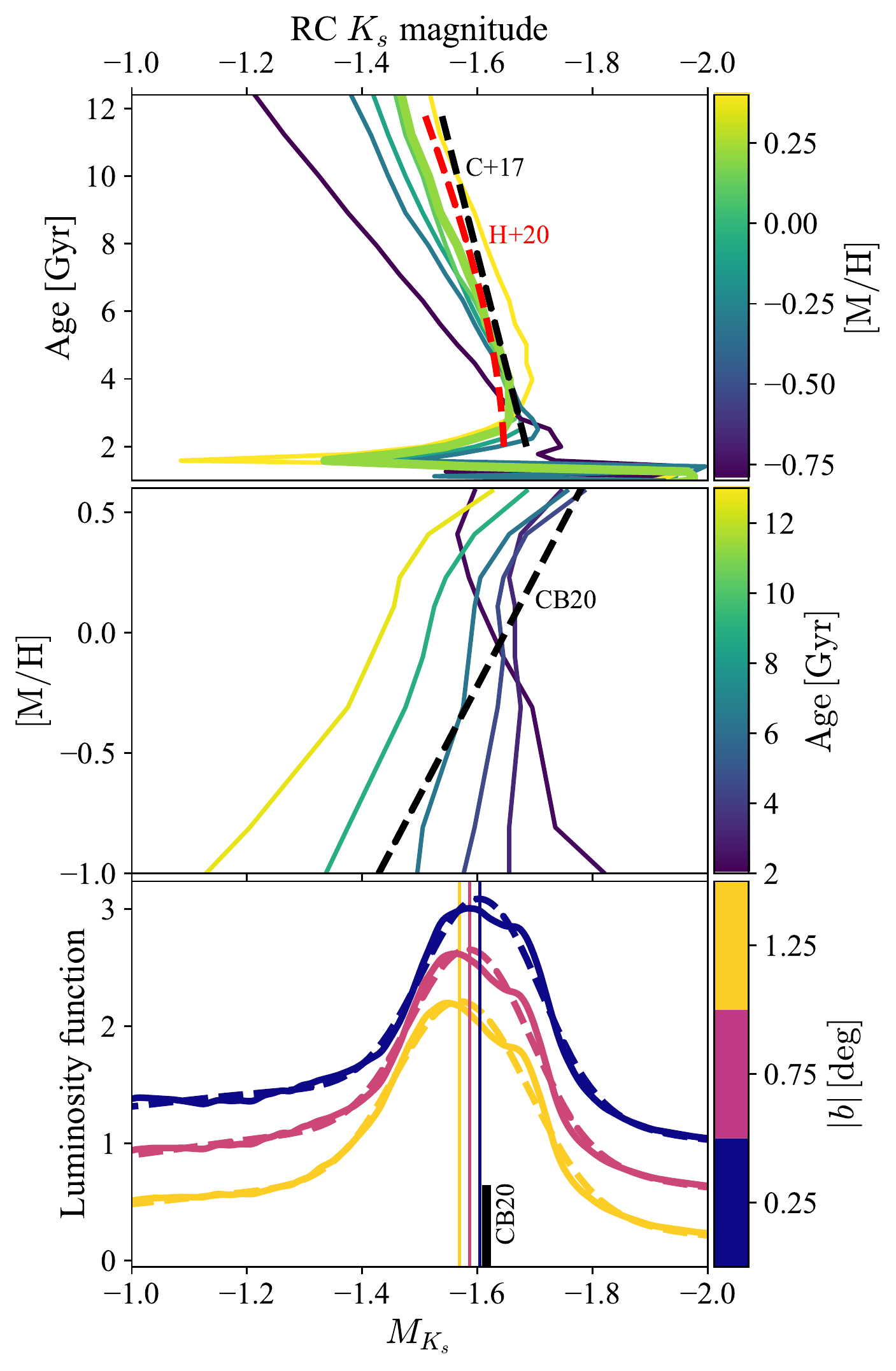}
    \caption{Properties of the red clump from PARSEC isochrones. The top panel shows the peak of the red clump distribution as a function of age coloured by metallicity. Two observational results are overplotted from \protect\citet[][C+17]{Chen2017} and \protect\citet[][H+20]{Huang2020} in black and red dashed respectively. The middle panel shows similar but as a function of metallicity coloured by age with the observational result from \protect\citet[][CB20]{ChanBovy2020} overplotted. The bottom panel shows the luminosity function for all giant stars (solid lines) coloured by Galactic latitude computed using an approximate star formation history from \protect\cite{Bernard2018} and spectroscopic metallicity distributions from a collection of sources (see text). The distributions are vertically offset for clarity. Dashed lines show a fit using two Gaussians (red clump and red giant branch bump) plus a quadratic (red giants). Vertical lines show the mean of the Gaussian fitted to the red clump. The black line is the solar neighbourhood result from \protect\cite{ChanBovy2020}.}
    \label{fig:parsec_models}
\end{figure}
To convert the selective extinction ratios, $E(c)/E(H-K_s)$, of the previous section to absolute extinction ratios, $A_x/A_{K_s}$, we must adopt a reference absolute extinction ratio value of $R_{H-K_s}$. This can be measured using the slope of the red clump in colour-magnitude space \citep{AlonsoGarcia2017} or from the absorption of hydrogen lines as in the work of \cite{Fritz2011}. Here we employ a method for estimating the absolute extinction ratios using the apparent $K_s$ magnitude of the red clump. In the absence of other information, there is a complete degeneracy between extinction, distance and absolute magnitude. Here we break the degeneracy by assuming a fixed absolute magnitude of the red clump and that the peak of the density distribution of red clump stars occurs at the Galactic centre distance from \cite{GravityCollaboration2021}. Further, we shall see that using a range of fields with different extinctions breaks the degeneracy without assuming a red clump absolute magnitude. This method shares elements with the classic method of measuring the slope of the red clump feature in colour-magnitude space \citep[e.g.][]{Nishiyama2009,AlonsoGarcia2017,NoguerasLara2018} and is very similar to the methods of \cite{Matsunaga2016} and \cite{Dekany2019} who considered what the extinction law could plausibly be to put a group of Cepheids at the Galactic centre. 

\subsection{Model}
We only consider stars within $11.5<K_{s0}<14.5$, and if $J$ and $H$ are available additional colour cuts of $0.2-1.5\sigma_{JK}<(J-K_s)_0<1+1.5\sigma_{JK}$ and $-0.3-1.5\sigma_{HK}<(H-K_s)_0<0.5+1.5\sigma_{HK}$ (initially using the \citealt{AlonsoGarcia2018} extinction curve) and the extinction spreads $\sigma_i$ estimated from the red clump method (see Appendix~\ref{appendix::rc_method}). We initially bin the sky in Healpix with NSIDE=128 and subdivide until all bins have no fewer than $5000$ stars. We define our model of the extinction-corrected magnitude distribution for an on-sky bin $(\ell,b)$ as
\begin{equation}
\begin{split}
    p(K_{s0}) &= w M(K_{s0})+(1-w)B(K_{s0}),\\
    M(K_{s0})&=N_M^{-1}\mathcal{C}(K_{s0})\int \mathrm{d}K_{s0}' \mathcal{N}(K_{s0}|K_{s0}',\sigma_{K_s}) M_0(K_{s0}'),\\
    M_0(K_{s0}')&=\exp\Big(\frac{3\ln{10}K_{s0}'}{5}\Big) \sum_{i=1}^{i=N_G}\omega_i\mathcal{N}(K_{s0}'|\mu_{i,K_s},\Sigma_{i,K_s}),\\
    B(K_{s0}) &= N_B^{-1}\Big(1+aK_{s0}+bK_{s0}^2\Big),\\
    \mathcal{N}(x|\mu,s)&=\frac{1}{\sqrt{2\pi s^2}}\exp\Big(-\frac{(x-\mu)^2}{2s^2}\Big).
\end{split}
\end{equation}
Our model is a mixture of a background quadratic model, $B$, summed with one or two components, $M$, which represent the red clump and red giant branch bumps. The density models of red clump and red giant branch bump stars are Gaussians $\mathcal{N}(K_{s0}'-\mu_{i,K_s},\Sigma_{i,K_s})$ which are transformed to magnitude space with the Jacobian $\exp\Big(\frac{3\ln{10}K_{s0}'}{5}\Big)$. We convolve these magnitude distributions with a Gaussian of width $\sigma_{K_s}$ which is the quadrature sum of the intrinsic magnitude spread of the red clump \citep{ChanBovy2020}, the extinction spread and the median magnitude error. $\mathcal{C}^{-1}(K_{s0})$ is the completeness at unextincted magnitude $K_{s0}$ found by interpolating unextincted magnitude against completeness, which is a function of extincted magnitude, for the stars in each on-sky bin. The completeness analysis is presented in Appendix~\ref{section::completeness}.
The normalization factors $N_M$ and $N_B$ are the model and background integrated over the magnitude range. For speed, we evaluate $M(K_{s0})$ on a grid and interpolate for each star as well as the normalization integrals. 

For each on-sky bin, we initially extinction correct $K_s$ using the red clump and Rayleigh Jeans colour excess method maps described in Appendix~\ref{appendix::rc_method} and Appendix~\ref{appendix::rjce_method}. $E(H-K_s)_\mathrm{RC}$ is used if $A_{Ks}<2$ and $E(H-[4.5])$ otherwise as it is more reliable in high extinction regions where the red clump stars around the Galactic centre are so heavily extincted they are lost in VIRAC2 $H$. We use the extinction law computed from the previous section combined with \cite{AlonsoGarcia2018} to find $R_{H-[4.5]}=A_{K_s}/E(H-[4.5])$. Although 2D extinction maps neglect the extinction variation along the line-of-sight, they should accurately capture the extinction to the peak stellar density so on the average should correctly estimate the extinction to the bulge/nuclear stellar disc populations. As a compromise to this, \cite{NoguerasLara2021} consider the dust as composed of two layers but the second layer corresponds to the peak stellar density. This `dust-screen' model could introduce biases for significant line-of-sight extinction variation through the bulge region, which we account for later in a model variant.

We then fit the model by maximising the log-likelihood for a model with one or two Gaussians and take that with the lower Akaike information criterion. We fit the parameters $a,b,\mu_{i,K_s},\Sigma_{i,K_s},\omega_i,w$. The parameter of interest is $\mu_{0,K_s}$ which gives the peak unextincted apparent magnitude of the red clump. Note that this value is distinct from the peak of the magnitude distribution as we will discuss shortly. We then compute the expected $\mu_{0,K_s}$, $\mu'_{0,K_s}$, from an assumed red clump absolute magnitude $M_{K_s,\mathrm{RC}}$ and distance to the population of
$(8.275\pm0.034)\,\mathrm{kpc}$ \citep{GravityCollaboration2021}. Note that the uncertainty in distance modulus to the Galactic centre is $\sim0.01\,\mathrm{mag}$ which is smaller than the uncertainty on the red clump magnitude, which we will discuss later. However, this rests on the assumption that the density structure of the Galaxy on scales of $\sim1\,\mathrm{deg}$ peaks at the location of Sgr A*. This is supported by evidence that it is the dynamical centre of the Galaxy from the kinematics of the bulge \citep{Sanders2019a,ClarkeGerhard2021}, the nuclear stellar disc \citep{Schoenrich2015,Sormani2021} and gas kinematics \citep[e.g.][]{Sormani2015}. For each on-sky bin, we compute $\Delta \mu=\mu_{0,K_s}-\mu'_{0,K_s}$ and calculate an updated $R_{H-[4.5]}=(1+\Delta\mu/\langle A_{K_s}\rangle)R_{H-[4.5]}$. We repeat the entire procedure using the updated extinction law (including generating the utilised sample which depends on the choice of extinction law). Systematic errors in $\mu_{0,K_s}$ translate into systematic errors in $R_{H-[4.5]}$ as $\Delta R_{H-[4.5]} = \Delta\mu_{0,K_s}/\langle E(H-[4.5])\rangle$.

It is worth briefly considering how the mode of the magnitude distribution varies with the distribution of the red clump stars. Assuming all red clump stars have the same absolute magnitude, $M_\mathrm{RC}$, and trace a Gaussian density distribution $\rho(\bs{x})=\mathcal{N}(\bs{x}-\bs{x}_0,\sigma)$, the observed unextincted magnitude distribution is
\begin{equation}
    p(K_s) = s^3 \rho(\bs{x}) \approx s^3 \mathcal{N}(s-s_0,\sigma),
    \label{eqn::mode_shift}
\end{equation}
for lines of sight close to the Galactic centre, which is a distance $s_0$ away. Here $K_s=M_\mathrm{RC}+5\log_{10}(100 s/\mathrm{kpc})$. Note that although the peak of the density distribution is at $K_s=K_{s,\rho}=M_\mathrm{RC}+5\log_{10}(100 s_0/\mathrm{kpc})$, the peak of the magnitude distribution is at $K_s\approx K_{s,\rho}+3(5/\ln10)\sigma^2/s_0^2$ (assuming $\sigma/s_0\ll1$). This means intrinsically broader density distributions will have fainter peaks in the magnitude distribution. For instance, a density distribution of width $1(2)\,\mathrm{kpc}$ would give rise to a magnitude shift of $\sim0.1(0.4)\,\mathrm{mag}$. Without proper modelling this would result in a larger inferred distance of $\Delta s = 3\sigma^2/s_0$ (so $\sim0.4(1.5)\,\mathrm{kpc}$ for our example). Note that neither broadening from magnitude uncertainties, nor the width of the red clump nor extinction give rise to this effect.

After fitting each bin, we measure the difference in red clump location with respect to the expected and divide by the extinction estimated from the mean $(H-[4.5])$ colour to find $R_{H-[4.5]}$. 

\begin{figure}
    \centering
    \includegraphics[width=\columnwidth]{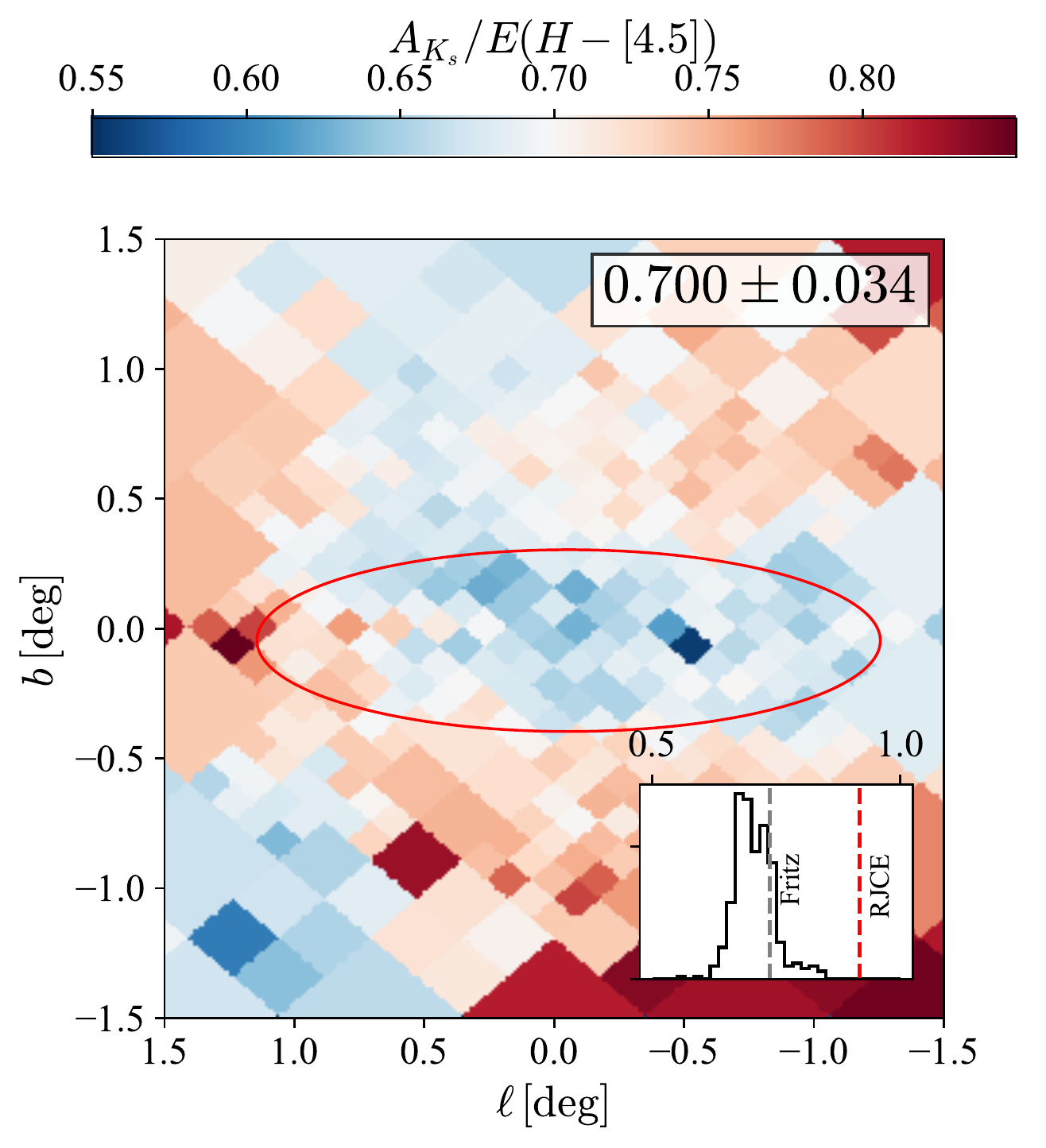}
    \caption{Variation of $R_{H-[4.5]}=A_{K_s}/E(H-[4.5])$ across the central $3\times3\,\mathrm{deg}^2$ assuming a fixed red clump magnitude of $M_{K_s,\mathrm{RC}}=-1.622$. The ellipse gives the approximate extent of the nuclear stellar disc. The inset plot displays the distribution of values along with other measurements.}
    \label{fig:onskymap_akh45}
\end{figure}
\begin{figure}
    \centering
    \includegraphics[width=\columnwidth]{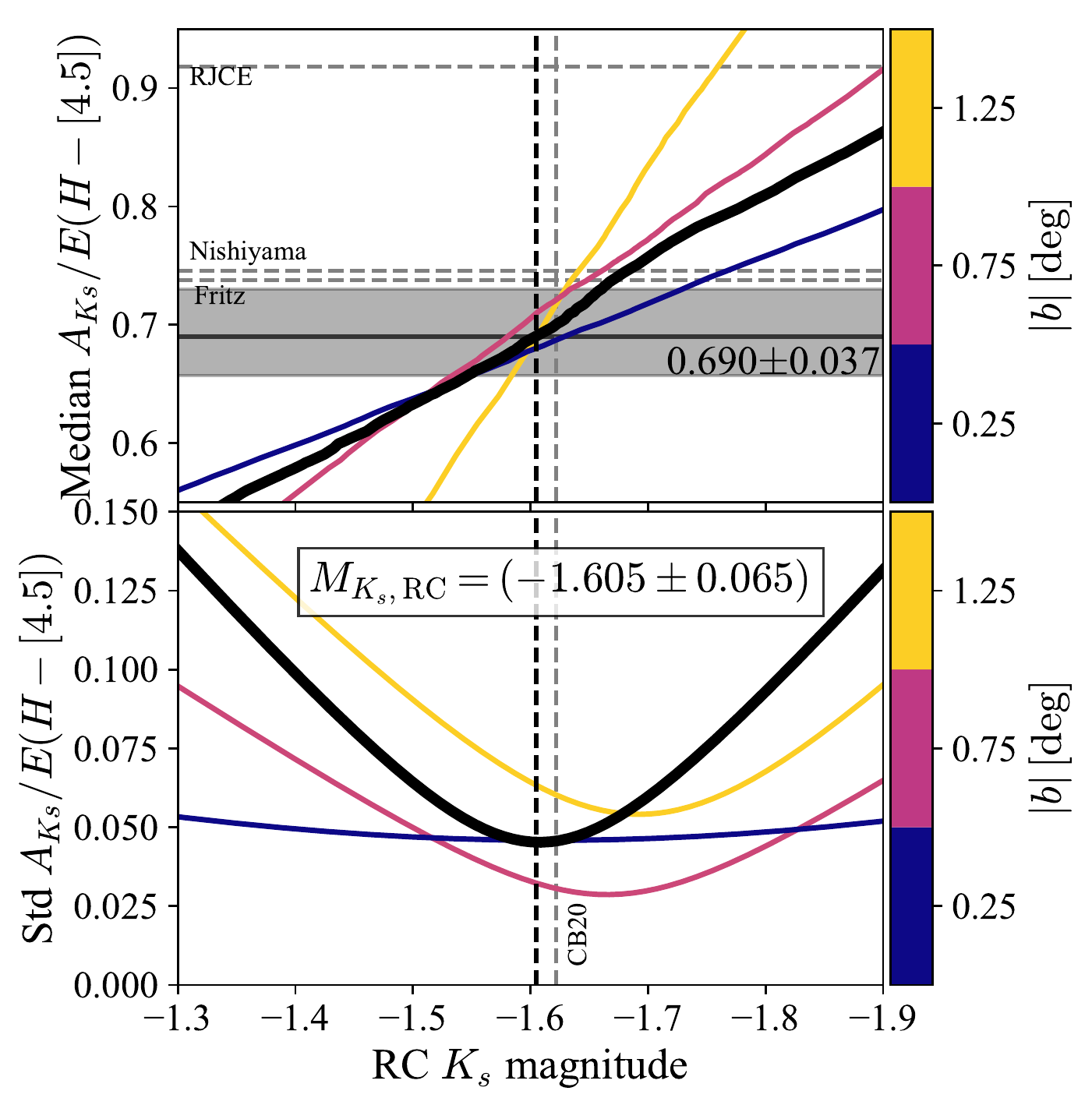}
    \caption{Results of fitting the red clump density peak. The top panel shows the variation of $R_{H-[4.5]}=A_{K_s}/E(H-[4.5])$ with the assumed red clump absolute magnitude. The coloured lines show the median computed over different on-sky regions as shown in the colour bar. Solid black is for the entire region. The horizontal dashed lines show the measurements from other authors. The second panel shows the standard deviation of $A_{K_s}/E(H-[4.5])$. The absolute magnitude of the red clump at the minimum standard deviation in $A_{K_s}/E(H-[4.5])$ is marked as a black dashed line. The horizontal band in the upper panel gives the median $A_{K_s}/E(H-[4.5])$ and $\pm1\sigma$ computed over on-sky pixels for this choice of red clump magnitude. }
    \label{fig:absmag_results}
\end{figure}

\subsubsection{Red clump properties}
Red clump stars are the low mass helium burning stage of stellar evolution. \cite{GirardiReview} reviews the properties of red clump stars as well as their use as distance and extinction tracers. In stars with initial masses below $\sim1.5\,M_\odot$, the post-main sequence core is electron-degenerate. During hydrogen shell burning on the giant branch the core accretes until a core mass of around $0.5\,M_\odot$ is reached and the electron degeneracy is lifted through a series of He flashes. This produces a core mass independent of initial mass and hence gives rise to the standard candle nature. In higher mass stars, electron degeneracy is not reached in the core and core He burning proceeds after the main sequence producing a monotonic relationship between core and initial mass, giving rise to secondary red clump stars, a fainter extension of the red clump. Despite the near constancy of the core mass, variations in the absolute magnitude of red clump stars occur due to metallicity and age variations, although these tend to be minimised in the $K_s$ band. \cite{GirardiReview} also highlights that, even at fixed age and metallicity, the skew of the magnitude distribution of the red clump can be significant, although the distribution becomes more symmetric for older populations.

In Fig.~\ref{fig:parsec_models} we show the properties of the peak of the red clump $K_s$ magnitude distribution as a function of age and metallicity for a set of PARSEC isochrones \citep{Bressan2012,Chen2014a,Chen2015,Tang2014,Marigo2017}. Below $\sim2\,\mathrm{Gyr}$ `red clump' stars burn helium in non-degenerate conditions giving rise to significant variation in the luminosity with mass and producing the so-called secondary clump. Above this age, the behaviour of the absolute magnitude of the red clump stars is more constant. The older, lower mass stars tend to be fainter than their younger counterparts, a trend that steepens for more metal-poor stars. Both \cite{Chen2017} and \cite{Huang2020} have provided calibrations for $M_{K_s,\mathrm{RC}}$ as a function of age via asteroseismic measurements \citep[here we use the calibration for {[Fe/H]}>0.1 from][]{Huang2020}, which match well the behaviour of the PARSEC isochrones with a typical gradient of $\sim0.015\,\mathrm{mag}\,\mathrm{Gyr}^{-1}$. It should be noted however that the age distribution of red clump stars are known to not trace the underlying star formation history. \cite{Bovy2014} shows how for a flat star formation history the red clump age distribution is strongly skewed towards young stars with a mode around $2\,\mathrm{Gyr}$. Therefore, even for star formation histories biased to early star formation the mean red clump age will be significantly smaller than the peak star formation epoch (provided star formation is not strongly truncated at early times). Fig.~\ref{fig:parsec_models} also shows how typically the red clump $K_s$ magnitude becomes brighter for more metal-rich populations with the gradient steepening for older populations. The gradient for the PARSEC isochrones for older populations agrees well with the gradient of $-0.21\,\mathrm{mag}\,\mathrm{dex}^{-1}$ found from \cite{ChanBovy2020} using APOGEE and Gaia DR2 data. \cite{ChanBovy2020} also show how there is a magnitude gradient with $[\alpha/\mathrm{Fe}]$ for the APOGEE stars of $\sim0.33\,\mathrm{mag}\,\mathrm{dex}^{-1}$. Combining with the asteroseismic measurements, a gradient of $[\alpha/\mathrm{Fe}]$ with age of $\sim0.05\,\mathrm{dex}\,\mathrm{Gyr}^{-1}$ would produce such an effect. This is typically the gradient observed in the solar neighbourhood \citep{Feuillet2019}. Therefore, it is unclear observationally what role alpha-enhancement plays in altering the red clump magnitude.

\subsubsection{Red clump distribution in the inner Galaxy}
With the theoretical and observational considerations of the previous section, we now turn to predicting the expected behaviour of the red clump magnitude for the considered sample of stars. The metallicity distribution of the inner Galaxy has been revealed through several spectroscopic surveys e.g. ARGOS \citep{Ness2013}, APOGEE \citep{RojasArriagada2020}, GES \citep{RojasArriagada2014} and GIBS \citep{Zoccali2017}. The results from these surveys are in agreement that the metallicity distribution of the inner Galaxy consists of multiple sub-components \citep{Ness2013} and that there is a vertical metallicity gradient \citep{Gonzalez2013}. In the inner $3\times3\,\mathrm{deg}^2$ there are comparatively few spectroscopic studies outside of the NSC. The most recent data is summarised by \cite{Schultheis2019}. We have taken stellar metallicity data from APOGEE DR16 \citep{SDSSDR16}, \cite{Nandakumar2018} and GIBS \citep{Zoccali2014,Zoccali2017}. This sample has mean metallicity  $-0.07\,\mathrm{dex}$ and standard deviation $\sim0.4\,\mathrm{dex}$. The mean $[\mathrm{Fe}/\mathrm{H}]$ is $-0.18\,\mathrm{dex}$. In $[\alpha/\mathrm{Fe}]$ vs. $[\mathrm{Fe}/\mathrm{H}]$, the distribution follows a standard chemical evolution track with the metal-rich stars not displaying any $\alpha$-enhancement (possibly even a small $\alpha$ deficit). The metallicity distribution is clearly composed of a metal-rich and metal-poor component, the latter of which becomes more dominant at higher latitudes giving rise to a metallicity gradient of $-0.1\,\mathrm{dex}/\mathrm{deg}$.

The age distribution of the inner Galaxy is less well known. Traditionally, from photometry, the bulge has been viewed as an old structure \citep[e.g.][]{Zoccali2003} but this was thrown into question by spectroscopic ages of microlensed dwarfs \citep{Bensby2013}, many of which are young. Recent work by \cite{Bernard2018} constrained the age distribution of the bulge ($-5\,\mathrm{deg}\lesssim b<-2\,\mathrm{deg}$) from Hubble Space Telescope photometry of the main sequence turn-off stars, concluding that, although the bulge is predominantly old, approximately $10\percent$ of stars are younger than $5\,\mathrm{Gyr}$. This fraction increases to $\sim20\percent$ for more metal-rich ($[\mathrm{Fe}/\mathrm{H}]\gtrsim0.2\,\mathrm{dex}$) stars, consistent with the \cite{Bensby2013} work. Further evidence for a predominantly old ($\gtrsim8\,\mathrm{Gyr}$) bulge comes (indirectly) from [C/N] measurements of giant stars \citep{Bovy2019,Hasselquist2020}, although, as highlighted by \cite{Hasselquist2020}, age appears to correlate with both metallicity and Galactic height of the populations. \cite{NoguerasLara2020} have used the luminosity of red clump stars to conclude the majority ($\sim95\percent$) of the nuclear stellar disc formed more than $8\,\mathrm{Gyr}$ ago with some evidence of a more recent ($<1\,\mathrm{Gyr}$ ago) star formation burst \citep{Matsunaga2011}. This is consistent with ongoing/recent star formation within the central molecular zone \citep{Morris1996} and is broadly consistent with the conclusions of \cite{Bernard2018} on the wider bar/bulge but possibly suggesting the nuclear stellar disc is on average older than the surrounding bulge.
We have fitted by-eye a very simple star formation history to the `cleanest' combined fit from \cite{Bernard2018} of the form $\mathrm{sech}^2((13.5\,\mathrm{Gyr}-\tau)/4.7\,\mathrm{Gyr})$ with a truncation at $14\,\mathrm{Gyr}$. Combining the metallicity distributions and star formation histories with the PARSEC isochrones and adopting a \cite{Kroupa2001} initial mass function, we have computed the luminosity function of the giant branch stars in the inner bulge region. We show the results in the lower panel of Fig.~\ref{fig:parsec_models} along with a simple double Gaussian plus quadratic fit to represent the red clump stars, the red giant branch bump stars and the red giant branch stars respectively. We find that the lowest latitude bin has a red clump magnitude of  $M_{K_s,\mathrm{RC}}=-1.61\,\mathrm{mag}$. This agrees well with the mean solar neighbourhood result from \cite{ChanBovy2020} of $M_{K_s,\mathrm{RC}}=-1.622\,\mathrm{mag}$ and more specifically using their relations adopting the mean $(J-K_s)=0.647$ (see Appendix~\ref{appendix::rc_method}) and mean metallicity $-0.18\,\mathrm{dex}$ gives $M_{K_s,\mathrm{RC}}=-1.595\,\mathrm{mag}$. The metallicity gradient with latitude produces a red clump magnitude gradient of $0.032\,\mathrm{mag}\,\mathrm{deg}^{-1}$ whilst using the change in mean metallicity in combination with the results of \cite{ChanBovy2020} we would expect $0.024\,\mathrm{mag}\,\mathrm{deg}^{-1}$. At all latitudes the red clump distribution is well reproduced by a Gaussian with standard deviation $\sim0.11\,\mathrm{mag}$. \cite{ChanBovy2020} measured the solar neighbourhood red clump to have an intrinsic standard deviation of $0.097\,\mathrm{mag}$ which combined in quadrature with that arising from the metallicity variance predicts a standard deviation of $\sim0.13\,\mathrm{mag}$, similar to the PARSEC models. The red clump peaks from the PARSEC isochrones have a slight bimodal structure arising from the bimodal metallicity distributions such that the mode typically peaks $\sim0.03\,\mathrm{mag}$ fainter than the Gaussian mean.

\subsection{Results}
We apply the method described in the previous subsection using different assumptions for the red clump distribution. All results are given in Table~\ref{tab:absmag_results}. As our fiducial model we use the absolute $K_s$ magnitude from \cite{ChanBovy2020} of $M_{K_s,\mathrm{RC}}=-1.622$. \cite{ChanBovy2020} derived the absolute magnitude using 2MASS data. The $K_s$ bands in VVV and 2MASS differ according to $K_{s,\mathrm{VVV}} = K_{s,\mathrm{2MASS}} + 0.01(J-K_s)_\mathrm{2MASS}$ \citep{GonzalezFernandez2018} so the difference is negligible for red clump stars with $(J-K_s)\approx0.6$. As discussed in the previous section, the expectation from stellar models is within $\sim0.02\,\mathrm{mag}$ of this $M_{K_s,\mathrm{RC}}$. We use $\sigma_{K_s}=0.11\,\mathrm{mag}$ as the intrinsic red clump spread (consistent with the PARSEC isochrones and the expectation from \cite{ChanBovy2020}). We show the resulting on-sky $R_{H-[4.5]}=A_{K_s}/E(H-[4.5])$ map in Fig.~\ref{fig:onskymap_akh45}. We find that on average $R_{H-[4.5]}=(0.700\pm 0.034)$ where the errorbar is from the standard deviation across inspected Healpix. The variation of $\sim5\percent$ is similar to the variation in the colour excess ratios. Our value of $R_{H-[4.5]}$ can be compared to the results from \cite{Fritz2011} of $R_{H-[4.5]}=(0.74\pm0.06)$ and \cite{WangChen2019} of $R_{H-[4.5]}=(0.74\pm0.06)$. This ratio is used in the Rayleigh Jeans colour excess method (see Appendix~\ref{appendix::rjce_method}) for which \cite{Majewski2011} use $R_{H-[4.5]}=0.918$ based on the extinction laws from \cite{Indebetouw2005}. We see some clear correlated structure in Fig.~\ref{fig:onskymap_akh45}, most obviously the midplane within the nuclear stellar disc has lower $R_{H-[4.5]}$. This might imply a fainter red clump magnitude in this region possibly reflecting a star formation history more biased towards earlier times \citep{NoguerasLara2020}. In Fig.~\ref{fig:absmag_results} we show the median of $R_{H-[4.5]}$ as a function of the assumed red clump $M_{K_s,\mathrm{RC}}$. The slope of $R_{H-[4.5]}$ with red clump magnitude reflects the mean extinction. Lower extinction leads to steeper slopes. This means constraints from lower latitude bins are more trustworthy as they are less reliant on assumptions regarding the red clump magnitude. 

Instead of assuming a magnitude for the red clump, we can assume the extinction coefficient $R_{H-[4.5]}$ is constant over the inspected region. Due to the range of slopes in the top panel of Fig.~\ref{fig:absmag_results} there is a choice of red clump magnitude that reduces the spread in $R_{H-[4.5]}$. In the lower panel of Fig.~\ref{fig:absmag_results} we show the standard deviation of $R_{H-[4.5]}$ as a function of $M_{K_s,\mathrm{RC}}$. The minimum occurs at $M_{K_s,\mathrm{RC}}=-1.605$ where $R_{H-[4.5]}=(0.690\pm0.037)$. We derive an uncertainty in $M_{K_s,\mathrm{RC}}$ by fixing $R_{H-[4.5]}=0.690$ and finding the resultant spread in $M_{K_s,\mathrm{RC}}=0.065$. This gives us a best estimate for the inner bar/bulge red clump absolute $K_s$ magnitude as $(-1.60\pm0.06)\,\mathrm{mag}$. This agrees well with the expectation of $M_{K_s,\mathrm{RC}}=-1.595$ from the  \cite{ChanBovy2020} models when adjusted for the mean colour and metallicity \citep[and other results from][]{Alves2000,Laney2012,Chen2017,Hawkins2017,RuizDern2018,Hall2019} and gives good evidence of the standard candle nature of the red clump across a range of Galactic environments. Combining with the mean red clump colour measurements from Appendix~\ref{appendix::rc_method} we find $M_{J,\mathrm{RC}}=(-0.96\pm0.06)\,\mathrm{mag}$ and $M_{H,\mathrm{RC}}=(-1.46\pm0.06)\,\mathrm{mag}$.

In addition to the fiducial model, we also run some model variants to inspect any possible systematic uncertainty in our results. Based on the discussion of the previous subsection, we run a model with a gradient of the magnitude of the red clump with latitude ((b) in Table~\ref{tab:absmag_results}) . We opt for $\mathrm{d}M_{K_s,\mathrm{RC}}/\mathrm{d}|b|=0.03\,\mathrm{mag}\,\mathrm{deg}^{-1}$ which agrees approximately with the results from the PARSEC isochrones and the expectation from the results of \cite{ChanBovy2020}. As expected, we find a slightly lower $R_{H-[4.5]}=(0.690\pm0.037)$ reflecting the gradient of $R_{H-[4.5]}$ with latitude seen in Fig.~\ref{fig:onskymap_akh45}. Minimising the spread of the red clump magnitude (accounting for the vertical gradient) leads to $M_{K_s,\mathrm{RC}}=(-1.641\pm0.063)$ demonstrating the red clump in the Galactic centre may be $\sim0.02$ magnitudes brighter than the fiducial solar neighbourhood value possibly as it is more metal-rich \citep[e.g. see][]{Schultheis2019}. However, all of these results are still consistent with our fiducial model within the reported uncertainties. We run a further two model variants ((c) and (d) in Table~\ref{tab:absmag_results}) where we solely use the $E(H-K_s)$ extinction maps from the red clump method or solely the $E(H-[4.5])$ extinction maps from all giant branch stars. The differences with respect to the fiducial model are negligible. As we only consider sources within $1.5\,\mathrm{deg}$ of the Galactic centre, the magnitude gradient due to the bar introduces shifts of $\sim0.1\,\mathrm{mag}$ assuming a bar angle of $28\deg$. However, many of the observed stars are expected to be part of the nuclear stellar disc, the geometry of which is poorly known although \cite{Sormani2021} demonstrated an axisymmetric dynamical model gives a good representation of the currently quite limited data. Running a model using a non-zero bar angle ((e) in Table~\ref{tab:absmag_results}) does not produce any significant difference in $R_{H-[4.5]}$. Wwe run a model ((f) in Table~\ref{tab:absmag_results})with a coarser graining of the Healpix initially using NSIDE=64 and subdividing requiring at least $50000$ stars per bin. Again, this does not produce any significant difference. A final consideration is that the use of the 2d extinction maps assumes that the peak of the colour distribution corresponds to the extinction for stars at the peak density in the bulge. However, significant line-of-sight variation of the extinction through the bulge may skew the mean estimates. Using the fractional gradient of the $E(H-K_s)$ \cite{Schultheis2014} maps with distance (in kpc) averaged over each on-sky bin, $\langle(1/E(H-K_s))\mathrm{d}E(H-K_s)/\mathrm{d}s\rangle$, we find that the mean $A_{Ks}$ extinction is too large by a factor $2(8.275\,\mathrm{kpc})(\ln 10/5)^2\Sigma_{K_s}^2\langle(1/E(H-K_s))(\mathrm{d}E(H-K_s)/\mathrm{d}s)\rangle$, where $\Sigma_{K_s}$ is the fitted width of density distribution in magnitude space (this formula arises through similar considerations to the discussion of the shift of the mode magnitude below equation~\eqref{eqn::mode_shift}). After fitting the parameters of the model at each iteration, we correct $\Delta\mu$ and $\langle A_{K_s}\rangle$ in each on-sky bin by this overestimate. The results are shown as model (g) of Table~\ref{fig:absmag_results} where $R_{H-[4.5]}$ must be slightly larger than the fiducial case to compensate for the reduced $A_{K_s}$ but the results are all comfortably within the fiducial uncertainties. More complicated dust distributions along the line-of-sight could give rise to more complicated behaviour but this gives confidence that it is a relatively weak effect.

\begin{figure}
    \centering
    \includegraphics[width=\columnwidth]{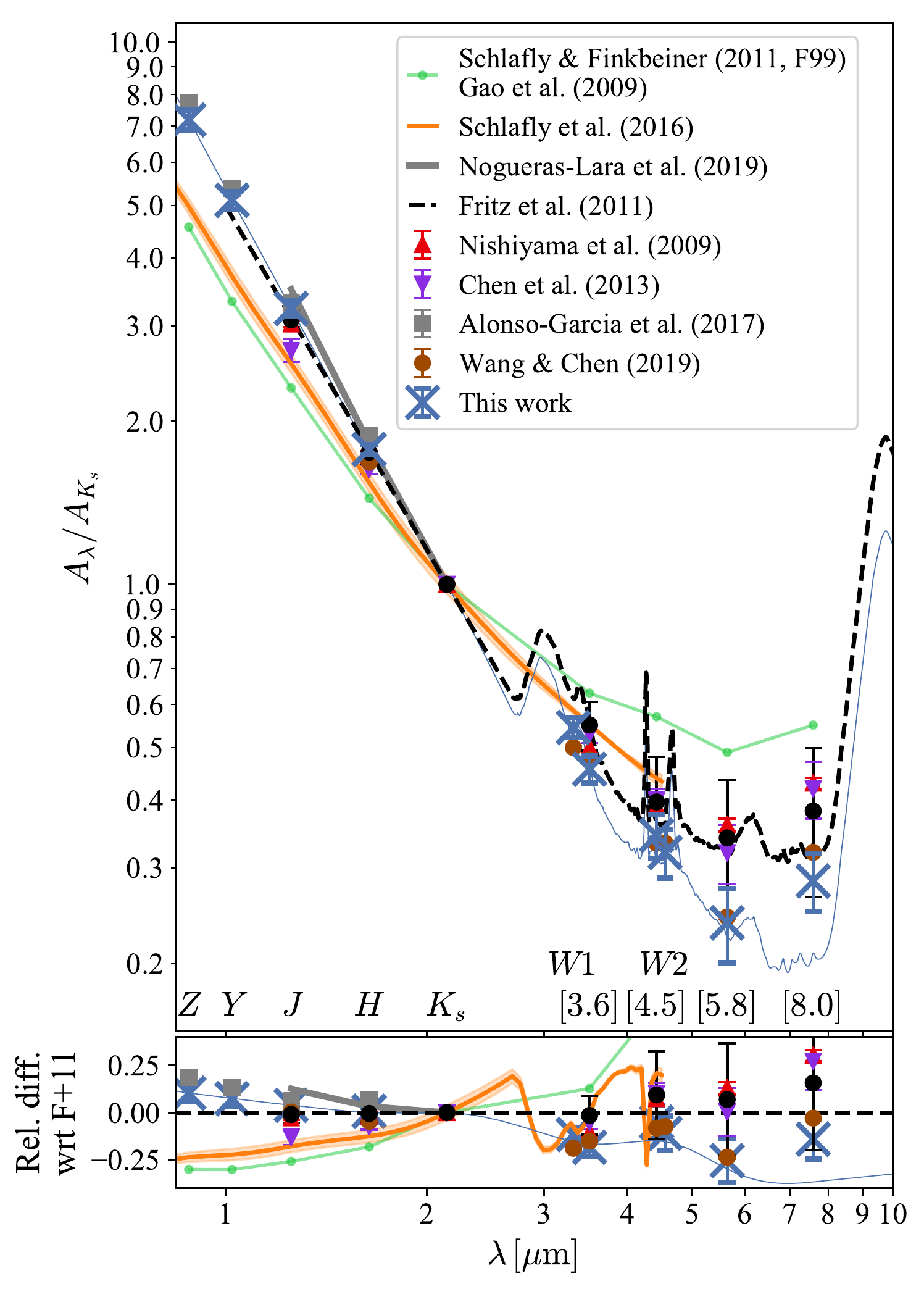}
    \caption{Our measured extinction curve (blue crosses and the monochromatic approximation from Table~\ref{tab:ak_approx} in thin blue) alongside other literature measurements (all corresponding to the Galactic centre/Galactic bulge region except 
    \protect\citealt{Gao2009},
    \protect\citealt{SchlaflyFinkbeiner2011} that uses the \protect\citealt{Fitzpatrick1999} law and \protect\citealt{Schlafly2016}). The errorbar combines the uncertainty in the selective extinction ratios from fitting the entire region and the variation in $R_{H-[4.5]}=A_{Ks}/E(H-[4.5])$ across the region. The lower panel shows the relative residual with respect to the \protect\cite{Fritz2011} extinction curve (black dashed top panel).}
    \label{fig:extinction_law}
\end{figure}

Using $R_{H-[4.5]}=(0.700\pm 0.034)$ in combination with the results from Table~\ref{tab:colour_ratio_results} we plot the resultant extinction law in Fig.~\ref{fig:extinction_law} and give the ratios $A_x/A_{K_s}$ in Table~\ref{tab:ext_results}. As found by other authors, the bulge extinction law is steeper than the more typical disc Milky Way extinction law from e.g. \cite{SchlaflyFinkbeiner2011} who give coefficients from a \cite{Fitzpatrick1999} extinction law and \cite{Schlafly2016} who utilised APOGEE data. We find $A_J/A_{K_s}$ and $A_H/A_{K_s}$ slightly lower than that found by \cite{AlonsoGarcia2017}. The lower $A_H/A_{K_s}$ is more in agreement with the results of \cite{Fritz2011}. We reproduce the low $A_{[5.8]}/A_{K_s}$ from \cite{Chen2018}. Using $\lambda_\mathrm{eff}=(1.25252, 1.63937, 2.14389)\,\mu\mathrm{m}$ for $J$, $H$ and $K_s$ respectively, the logarithmic infra-red extinction slope, or power-law index, is found as $\alpha=(2.22\pm0.08)$ using the $J$ and $H$ bands and $(2.14\pm0.08)$ using the $H$ and $K_s$ bands. This agrees within $1\sigma$ with both the measurement from \cite{SteadHoare2009} of $\alpha=(2.14\pm0.05)$ and the recent result of $\alpha=2.27$ from \cite{MaizApellaniz2020}, both studies using stars across the entirety of the Galaxy but also accounting for non-linearity in the extinction calculations. This is good evidence for a near universality to the near-infrared extinction law. \cite{Hosek2018} used observations of Westerlund 1 and red clump stars in the Galactic Centre to find $\alpha=(2.38\pm0.15)$ whilst using integrated photometry, \cite{StelterEikenberry2021} find $\alpha=(2.03\pm0.06)$. \cite{NoguerasLara2019} \citep[and similar in][]{NoguerasLara2020b} found evidence of a non-constant slope such that $\alpha=(2.43\pm0.10)$ for $J,H$ and 
$\alpha=(2.23\pm0.03)$ for $H,K_s$. Our results imply a non-constant $\alpha$ in the same sense as these measurements although the difference we find is much smaller and within our quoted uncertainties. As seen in Fig.~\ref{fig:onskymap_akh45}, in the midplane $|b|\lesssim0.1$ which is the region covered by \cite{NoguerasLara2019}, $R_{H-[4.5]}$ is lower if the red clump magnitude is constant across the entire region we have inspected. We find $\mathrm{d}\alpha_{JH}/\mathrm{d}R_{H-[4.5]}\approx1.3$ and  $\mathrm{d}\alpha_{HK_s}/\mathrm{d}R_{H-[4.5]}\approx2.2$. To transform Table~\ref{tab:colour_ratio_results} into the values of Table~\ref{tab:ext_results} using a different choice of $R_{H-[4.5]}$ involves
\begin{equation}
    \frac{A_x}{A_{Ks}}=1+\frac{1}{1.848R_{H-[4.5]}}\Big(1+\frac{E(H-K_s)}{E(x-H)}\Big).
\end{equation}

For ease of using the results in this paper, we provide a monochromatic approximation to the derived extinction law. We take the monochromatic extinction law, $A_\mathrm{F11}(\lambda)$, provided by \cite{Fritz2011} and multiply by a interpolating spline, $u(\lambda)$, with a series of nodes around the effective wavelengths of each filter considered here, and then a set beyond $15\,\mu\mathrm{m}$ with $y$-values of $0.7$ (an arbitrary choice that approximately matches the correction required at $8\,\mu\mathrm{m}$ and ensures the gradient of the spline tends to zero at large wavelength). We then adjust the spline $y$-values for $\lambda<15\,\mu\mathrm{m}$ computing the extinction coefficient for a $4750\,\mathrm{K}$ spectrum using equation~\eqref{eqn::ext_integral} and our adjusted extinction law, $A(\lambda)=A_\mathrm{F11}(\lambda)u(\lambda)$. The resulting monochromatic extinction is tabulated in Table~\ref{tab:ak_approx} and plotted in Fig.~\ref{fig:extinction_law}.

\begin{table*}
    \caption{Measurements of the absolute extinction ratio $R_{H-[4.5]}$. Each row corresponds to a different model. The columns display 
    (i) $R_{H-[4.5]}$ assuming a reference $M_{K_s,\mathrm{RC}}$ ($-1.622$ for all models except (b)),
    (ii) $R_{H-[4.5]}$ only for $|b|<0.2\,\mathrm{deg}$, 
    (iii) the gradient of $R_{H-[4.5]}$ with $\ell$, 
    (iv) the gradient of $R_{H-[4.5]}$ with $|b|$,
    (v) the gradient of $R_{H-[4.5]}$ with $A_{K_s}$,
    (vi) the near-IR extinction law slope measured from $A_J/A_H$ ($\alpha_{JH}$),
    (vii) the near-IR extinction law slope measured from $A_H/A_{K_s}$ ($\alpha_{HK}$),
    (vii) the choice of $M_{K_s,\mathrm{RC}}$ which minimises the spread in $R_{H-[4.5]}$ with the error given by the uncertainty in $R_{H-[4.5]}$ divided by the gradient of $R_{H-[4.5]}$ with $M_{K_s,\mathrm{RC}}$, and
    (viii) $R_{H-[4.5]}$ at this $M_{K_s,\mathrm{RC}}$. 
    The set of models are (a) our fiducial model with no vertical magnitude gradient, a width of the red clump of $0.11\,\mathrm{mag}$ and a combination of both $(H-K_s)_\mathrm{RC}$ and $(H-[4.5])$ for extinction in low and high extinction regions resceptively, (b) using a red clump absolute magnitude gradient of $\mathrm{d}M_{K_s,\mathrm{RC}}/\mathrm{d}|b|=0.03\,\mathrm{mag}\,\mathrm{deg}^{-1}$, 
    (c) extinction just using $(H-K_s)_\mathrm{RC}$, (d) extinction just using $(H-[4.5])$, (e) assuming the distribution of stars is aligned with the bar, (f) using an increased number of stars per bin and (g) accounting for a linear gradient of extinction with distance.}
    \centering
    \setlength\tabcolsep{3pt}
    \begin{tabular}{lccccccccc}
    &(i)&(ii)&(iii)&(iv)&(v)&(vi)&(vii)&(viii)&(ix)\\
Model&$R_{H-[4.5]}$&$R_{|b|<0.2}$&$\partial_\ell R$&$\partial_{|b|}R$&$\partial_{M_{K_s}}R$&$\alpha_{JH}$&$\alpha_{HK_s}$&$M_{K_s,\mathrm{RC}}$&$R_{\mathrm{min}}$\\\hline
(a) Fiducial&$0.700\pm0.034$&$0.678\pm0.036$&$-0.009$&$+0.033$&$-0.531$&$-2.22\pm0.08$&$-2.14\pm0.08$&$-1.605\pm0.065$&$0.690\pm0.037$\\
(b) Vertical gradient&$0.690\pm0.037$&$0.669\pm0.027$&$-0.016$&$-0.006$&$-0.527$&$-2.23\pm0.09$&$-2.16\pm0.09$&$-1.641\pm0.063$&$0.702\pm0.038$\\
(c) $E(H-K_s)$&$0.704\pm0.034$&$0.689\pm0.034$&$-0.010$&$+0.034$&$-0.535$&$-2.22\pm0.08$&$-2.13\pm0.08$&$-1.609\pm0.065$&$0.695\pm0.036$\\
(d) $E(H-[4.5])$&$0.702\pm0.038$&$0.667\pm0.028$&$-0.006$&$+0.021$&$-0.538$&$-2.22\pm0.09$&$-2.13\pm0.09$&$-1.609\pm0.067$&$0.693\pm0.038$\\
(e) Bar angle&$0.706\pm0.047$&$0.673\pm0.047$&$+0.042$&$+0.057$&$-0.531$&$-2.21\pm0.11$&$-2.13\pm0.11$&$-1.571\pm0.076$&$0.669\pm0.041$\\
(f) Coarser resolution&$0.694\pm0.037$&$0.671\pm0.021$&$+0.007$&$+0.045$&$-0.473$&$-2.22\pm0.09$&$-2.14\pm0.09$&$-1.593\pm0.060$&$0.679\pm0.041$\\
(g) Extinction gradient&$0.711\pm0.037$&$0.689\pm0.036$&$-0.007$&$+0.038$&$-0.539$&$-2.21\pm0.09$&$-2.11\pm0.08$&$-1.603\pm0.067$&$0.698\pm0.038$\\
\hline
    \end{tabular}
    \label{tab:absmag_results}
\end{table*}
\begin{table*}
    \centering
    \setlength\tabcolsep{3pt}
    \caption{Derived extinction law using $R_{H-[4.5]}=(0.700\pm0.034)$ from the fiducial model in Table~\ref{tab:absmag_results}. Here we have combined the colour excess measurements with the absolute extinction measurement, and we compare to pre-existing measurements (we quote the \citealt{Fritz2011} results for a $9480\,\mathrm{K}$ spectrum but the dependence on the source spectrum is weak). The first uncertainty comes from combining the uncertainty in the colour ratios fitted over the whole region (essentially negligible) together with the variance in $R_{H-[4.5]}$ across the region, whilst the second is the contribution from the variance in the colour ratios across the region. Note that \protect\cite{Nishiyama2009} report coefficients for the SIRIUS bands,
    \protect\cite{WangChen2019} for the 2MASS bands 
    and \protect\cite{NoguerasLara2019} for the HAWK bands (see the end of Section~\ref{sec::data} for approximate conversions).
    For each band, $x$, we report the effective wavelength $\lambda_\mathrm{eff}=\int\mathrm{d}\lambda\,T_x(\lambda)F_\lambda(\lambda)\lambda^2/\int\mathrm{d}\lambda\,T_x(\lambda)F_\lambda(\lambda)\lambda$ using a \protect\cite{CastelliKurucz2004} $T_\mathrm{eff}=4750\,\mathrm{K}$ spectrum $F_\lambda$.
    }
    \begin{tabular}{lccx{0.22\columnwidth}x{0.22\columnwidth}x{0.22\columnwidth}x{0.22\columnwidth}x{0.22\columnwidth}}
    \hline
Band&$\lambda_\mathrm{eff}\,[\mathrm{nm}]$&$A_x/A_{K_s}$&\cite{Fritz2011}&\cite{AlonsoGarcia2017}&\cite{Nishiyama2009}&\cite{NoguerasLara2019}&\cite{WangChen2019}\\
    \hline
$Z$&$879.583$&$7.19\pm0.30\pm0.19$&$-$&$7.74\pm0.11$&$-$&$-$&$-$\\
$Y$&$1021.237$&$5.11\pm0.20\pm0.14$&$4.64\pm0.22$&$5.38\pm0.07$&$-$&$-$&$-$\\
$J$&$1252.522$&$3.23\pm0.11\pm0.06$&$3.07\pm0.13$&$3.30\pm0.04$&$3.02\pm0.04$&$3.51\pm0.04$&$3.12\pm0.17$\\
$H$&$1639.371$&$1.77\pm0.04\pm0.04$&$1.75\pm0.08$&$1.88\pm0.03$&$1.73\pm0.03$&$1.81\pm0.01$&$1.68\pm0.11$\\
$[3.6]$&$3508.050$&$0.46\pm0.03\pm0.03$&$0.55\pm0.06$&$-$&$0.50\pm0.01$&$-$&$0.47\pm0.05$\\
$[4.5]$&$4421.765$&$0.34\pm0.03\pm0.05$&$0.40\pm0.08$&$-$&$0.39\pm0.01$&$-$&$0.33\pm0.04$\\
$[5.8]$&$5641.275$&$0.24\pm0.04\pm0.05$&$0.34\pm0.09$&$-$&$0.36\pm0.01$&$-$&$0.24\pm0.04$\\
$[8.0]$&$7591.510$&$0.28\pm0.03\pm0.06$&$0.38\pm0.12$&$-$&$0.43\pm0.01$&$-$&$0.32\pm0.04$\\
$W1$&$3317.236$&$0.54\pm0.02\pm0.03$&$-$&$-$&$-$&$-$&$0.50\pm0.06$\\
$W2$&$4552.394$&$0.32\pm0.03\pm0.04$&$-$&$-$&$-$&$-$&$0.33\pm0.05$\\
\hline
    \end{tabular}
    \label{tab:ext_results}
\end{table*}

\begin{table}
    \caption{A portion of the provided approximation to the monochromatic extinction law normalized with respect to the extinction at the $K_s$ effective wavelength ($2.144\,\mu\mathrm{m})$.}
    \centering
    \begin{tabular}{cc}
    Wavelength [$\mu\mathrm{m}$]&$A(\lambda)/A(2.144\,\mu\mathrm{m})$\\
    \hline
$0.800000	$&$8.952603$\\
$0.800920	$&$8.928876$\\
$0.801840	$&$8.905239$\\
$\cdots$&$\cdots$\\
$9.999080	$&$1.181260$\\
$10.000000  $&$1.181112$\\
    \end{tabular}
    \label{tab:ak_approx}
\end{table}


\subsection{Validation}
\begin{figure*}
    \centering
    \includegraphics[width=\textwidth]{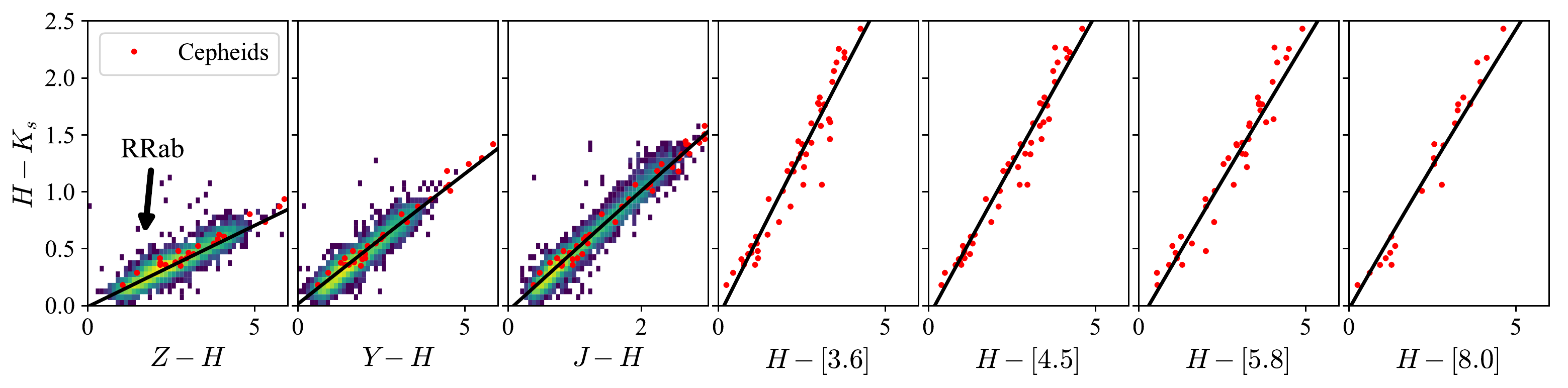}
    \includegraphics[width=\textwidth]{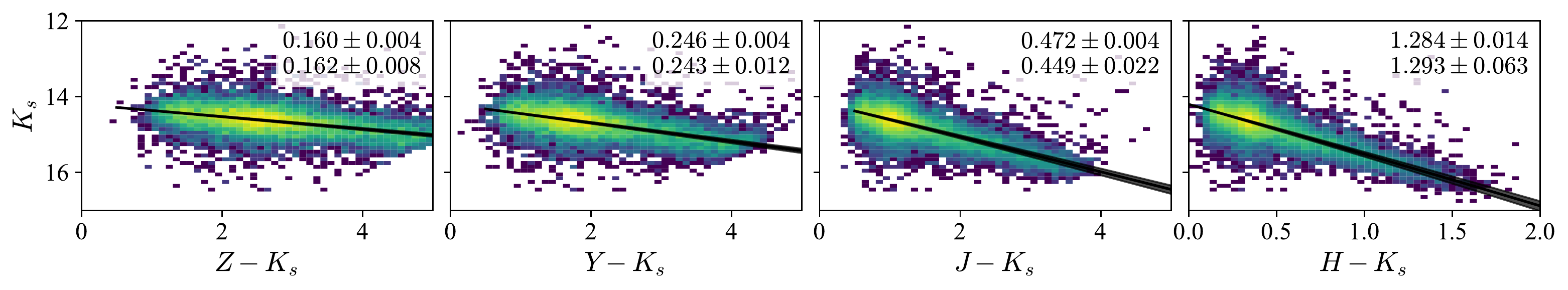}
    \caption{Colour-colour (upper) and colour-magnitude (lower) diagrams for Cepheids (red points) and RR Lyrae ab (histogram) with the derived extinction law overplotted. The reported numbers in the lower panels are the slope measured from the RR Lyrae ab sample (upper) and measured in this paper (lower).}
    \label{fig:rrl_cep}
\end{figure*}
\begin{figure}
    \centering
    \includegraphics[width=\columnwidth]{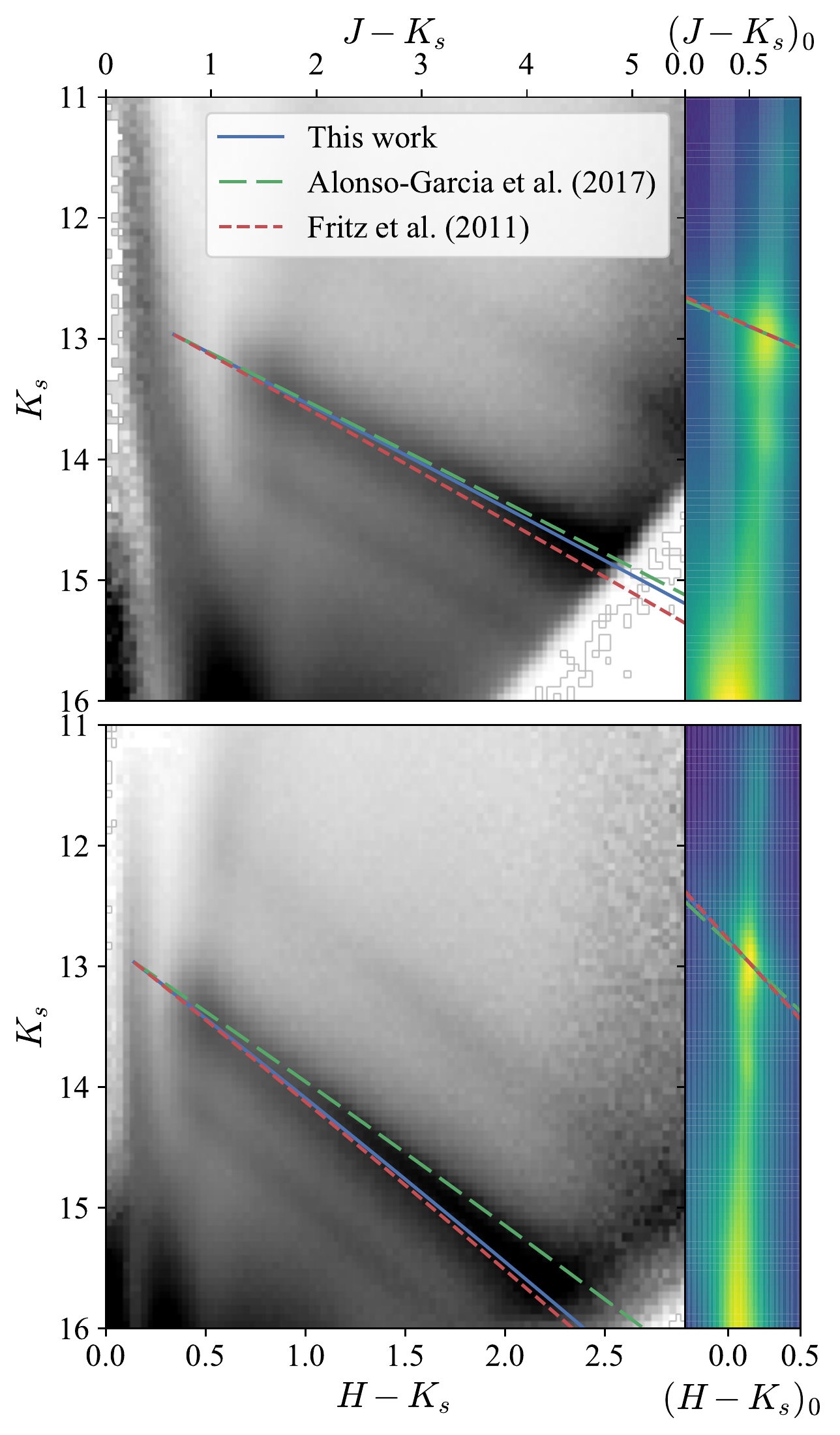}
    \caption{Colour-magnitude diagrams over the inner $3\times3\,\mathrm{deg}^2$ field. The grayscale shows the column-normalized density with a power law stretch with coefficient $0.4$. The extinction laws from this paper and previous works are displayed (with $\alpha_{JH}=(-2.223,-2.090,-2.088)$
    and $\alpha_{HKs}=(-2.135,-2.353,-2.092)$
    respectively), and the right panels show the unextincted colour-magnitude diagrams. The colours show the density with a power law stretch with coefficient $0.4$.}
    \label{fig:cmd}
\end{figure}
\begin{figure}
    \centering
    \includegraphics[width=\columnwidth]{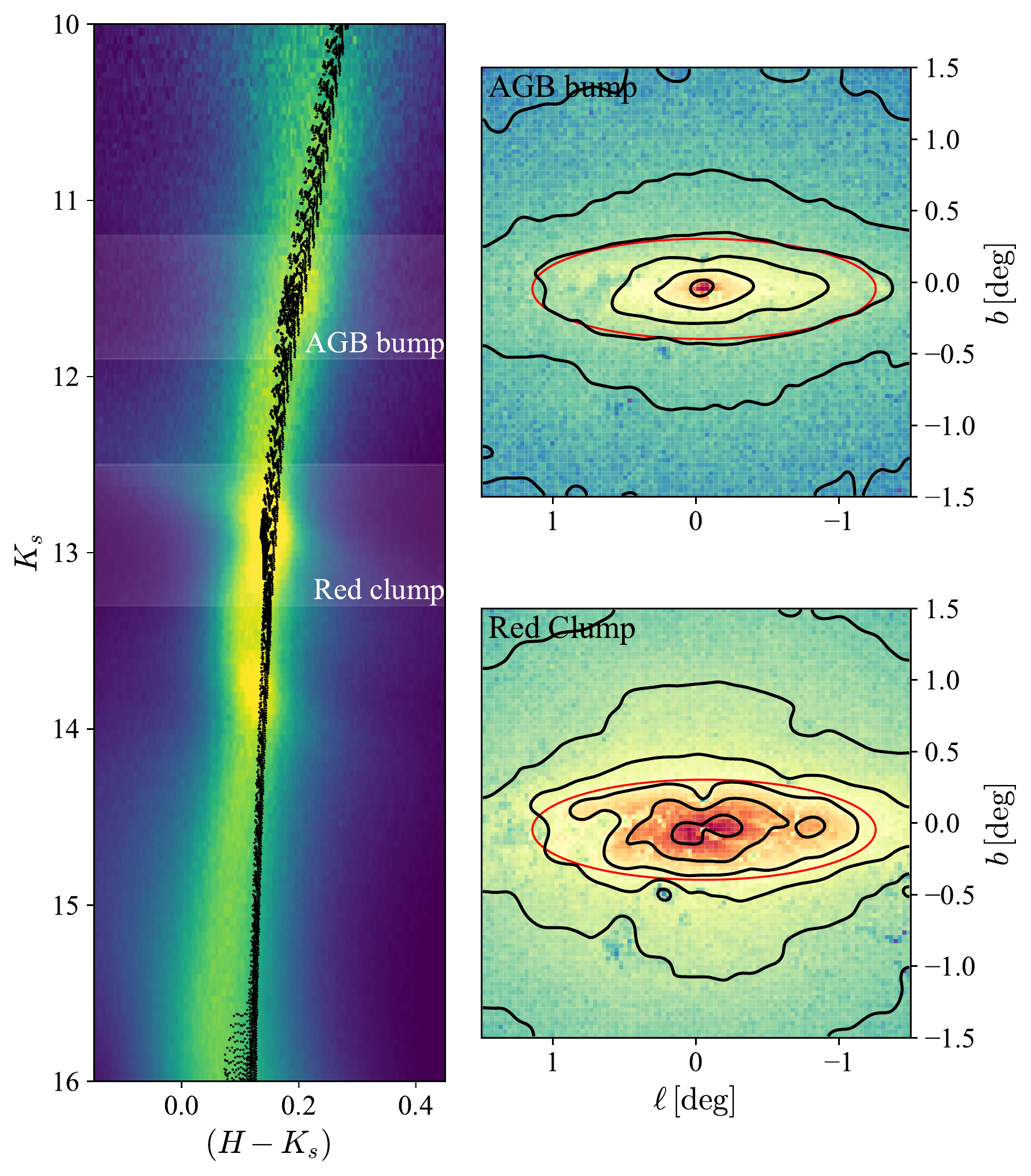}
    \caption{Number counts of stars in the nuclear stellar disc: the left panel shows the unextincted $(H-K_s)$ vs. $K_s$ row-normalized colour-magnitude diagram. Solar-metallicity PARSEC isochrones with $\log(\mathrm{age}/\,\mathrm{Gyr})>9.5$ at $8.275\,\mathrm{kpc}$ shifted in $(H-K_s)$ by $0.05\,\mathrm{mag}$ are overlaid in black. The AGB bump, red clump and red giant branch bump are clearly visible. In the right panels we show the number density of stars in two $K_{s0}$ bins around the AGB bump and red clump. We overlay a red ellipse centred on Sgr A* with aspect ratio $0.3$.}
    \label{fig:cnd}
\end{figure}
We close with some basic tests of the results derived here. First, we inspect the colour-magnitude and colour-colour diagrams for samples of classical Cepheids and RR Lyrae ab within the inner $6\times6\,\mathrm{deg}^2$ (to increase number statistics) from the work of \cite{Molnar2021}. These variables have been identified from the VIRAC2 catalogue using a hierarchical machine-learning classifier. We only use high-confidence variables (classification probability $>0.9$). The samples and the extinction vectors derived in this work are shown in Fig.~\ref{fig:rrl_cep}. We find there is very good agreement between the colour-colour and colour-magnitude distributions and the corresponding extinction vectors.

We extract a sample of VIRAC2 stars detected in at least $20\percent$ of observations covering their positions and between $K_s=11$ and $K_s=19$. We use our 2D extinction maps to extinction correct the $K_s$ magnitudes (assuming a fixed extinction law as reported in Table~\ref{tab:ext_results}). The colour magnitude diagram is shown in the left panel of Fig.~\ref{fig:cnd}. We extract two subsamples by selecting stars in two bins in unextincted magnitude: $11.2<K_{s0}<11.9$ corresponding to the AGB bump and $12.5<K_{s0}<13.3$ corresponding to the red clump. We display histograms of the resulting samples in the right panels of Fig.~\ref{fig:cnd}. We clearly note in both samples the presence of the nuclear stellar disc which is signifncant for $|b|<0.4\,\mathrm{deg}$. We overlay an ellipse with aspect ratio $0.3$ which gives a good `by-eye' match to the equi-density contours, particularly for the AGB bump sample. This agrees well with the diameter-to-thickness ratio of $5:1$ reported by \cite{Launhardt2002} based on fits to integrated NIR and FIR light, and axis ratio of $0.79$ from \cite{GallegoCano}. Outside of the nuclear stellar disc, there is a transition to the inner bulge, which has a rounder, yet still significantly flattened, shape.

\section{Conclusions}\label{sec::conclusions}
We have measured the extinction law from $0.9$ to $8$ microns in the inner $3\times3\,\mathrm{deg}^2$ of the Galaxy. Our method involves a combination of measuring the selective extinction ratios from colour-colour diagrams of bar/bulge red giant branch stars from VVV and GLIMPSE, and measuring the absolute extinction ratio by requiring the density of red clump giant stars peaks at the distance to the Galactic Centre measured by \cite{GravityCollaboration2021}. Assuming a red clump $K_s$ magnitude similar to that observed for stars in the solar neighbourhood, we have found that the extinction law is steep with a power-law index ($A_x\propto\lambda_x^{-\alpha}$) of $\alpha\approx2.2$ with only a weak indication of a non-constant slope across the $J$, $H$ and $K_s$ wavelength range. We have confirmed previous work that finds a low $A_{[5.8]}/A_{K_s}=(0.24\pm0.04)$. Furthermore, we have calibrated the Rayleigh-Jeans colour excess method as $A_{Ks}=0.677(H-[4.5]-0.188)$ for typical Galactic bulge stars with $K_s=13$. Our work incorporates the non-linearity of the extinction with total extinction, an important effect particularly for the IRAC $[8.0]$ band due to the $9.7\,\mu\mathrm{m}$ silicate feature. Our methods have used adaptive 2d extinction maps constructed from the mean $(J-K_s)$ and $(H-K_s)$ colours of the red clump stars and $(H-[4.5])$ of all giant stars as in high extinction regions red clump stars are too faint in $J$ and $H$ for VVV. 

Instead of assuming a red clump $K_s$ magnitude, we have demonstrated by ensuring the extinction law is constant over the surveyed region (as suggested by inspecting the spatial variation of the selective extinction ratios) we can independently constrain the red clump $K_s$ magnitude as $M_{K_s,\mathrm{RC}}=(-1.61\pm0.07)$. Reasonable systematic variations of the fiducial model such as accounting for vertical metallicity gradients or potential asymmetries related to the bar give results consistent with this. This measurement is very similar to the mean value observed for solar neighbourhood stars and gives confidence in the use of red clump stars as standard candles throughout the Galaxy.

\section*{Acknowledgements}
We thank the referee 
Jes\'us Ma\'iz Apell\'aniz
for useful comments that improved the presentation of the work.
J.L.S. acknowledges support from the Royal Society (URF\textbackslash R1\textbackslash191555). 
D.M. gratefully acknowledges support by the ANID BASAL projects ACE210002 and FB210003, and by Fondecyt Project No. 1220724.
Based on data products from observations made with ESO Telescopes at the La Silla or Paranal Observatories under ESO programme ID 179.B-2002. 
This research has made use of the SVO Filter Profile Service (\url{http://svo2.cab.inta-csic.es/theory/fps/}) supported from the Spanish MINECO through grant AYA2017-84089.
This paper made use of
\textsc{numpy} \citep{numpy},
\textsc{scipy} \citep{scipy}, 
\textsc{matplotlib} \citep{matplotlib}, 
\textsc{seaborn} \citep{seaborn} and
\textsc{astropy} \citep{astropy:2013,astropy:2018}.
\section*{Data availability}

All of the data used in this article are in the public domain, except for the VIRAC2 data which will be made available together with the catalogue description paper (Smith et al., in prep.).

\bibliographystyle{mnras}
\bibliography{bibliography}

\appendix
\section{Non-linearity of $JHK_s$ extinction bands}\label{app:nl_allbands}
\begin{figure*}
    \centering
    \includegraphics[width=0.8\textwidth]{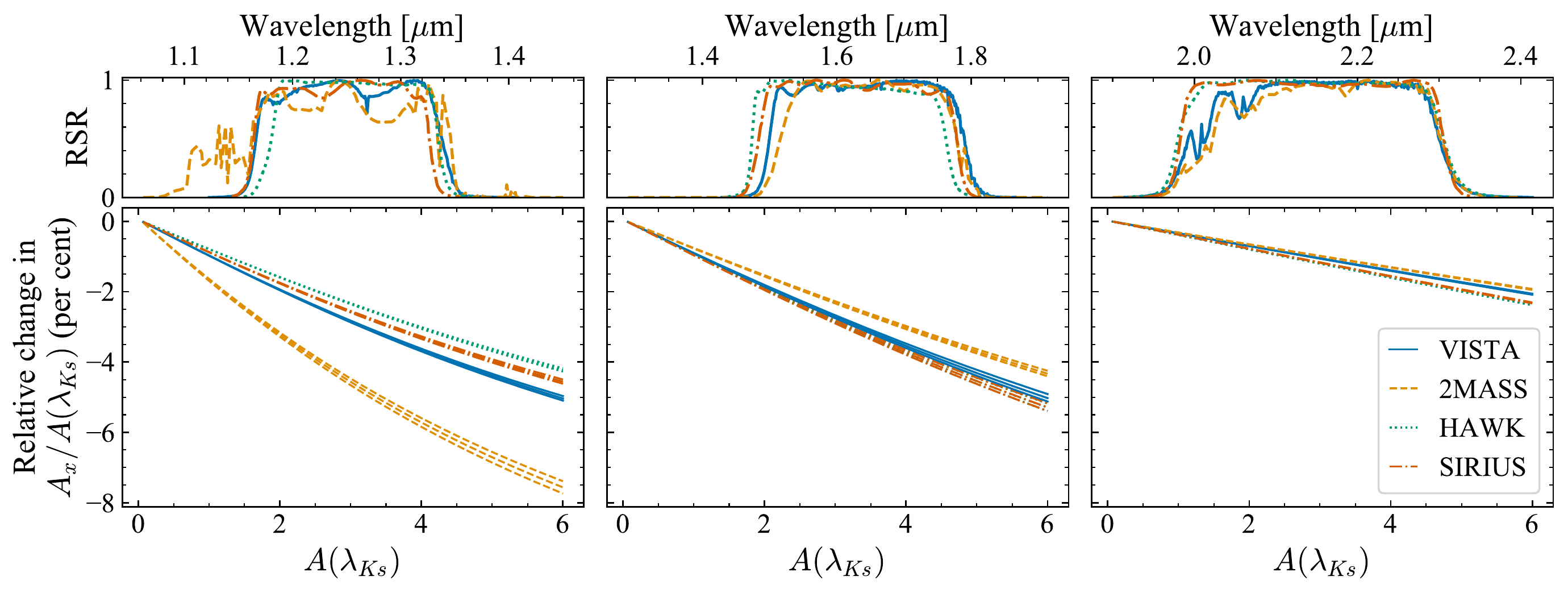}
    \caption{Comparison of the non-linearity of the extinction law in 2MASS (solid), VISTA (dashed), HAWK (dotted) and SIRIUS (dash-dot) $J$ (left), $H$ (centre) and $K_s$ (right) bands. We display curves of the relative percentage change in the extinction in band $x$, $A_x$ (evaluated using equation~\eqref{eqn::ext_integral}), normalized by the extinction evaluated at $\lambda_{K_s}=2.149\,\mu\mathrm{m}$. For each band, three curves are shown corresponding to three giant spectra as described in Section~\ref{section::nonlinearity} and we use the extinction curve from \protect\cite{Fritz2011}. The top row of panels shows the relative spectral response of each filter.}
    \label{fig:nl_allbands}
\end{figure*}

As a complement to Fig.~\ref{fig::nonlinear}, we provide equivalent plots comparing the non-linearities in the 2MASS, VISTA, HAWK and SIRIUS $J$, $H$ and $K_s$ bands in Fig.~\ref{fig:nl_allbands}. The strength of the non-linearities is related to the width of the photometric band: $J_\mathrm{2MASS}$ has significant non-linearities due to the blue transmission.

\section{2D Extinction maps}
Reliable extinction maps are an essential part of the interpretation of data for the central regions of our Galaxy. For simplicity, in this paper we work with 2D extinction maps and neglect the variation of the extinction along the line-of-sight. Whilst such an approach may be problematic when considering foreground or background stars, the 2D methods tend to measure the mean extinction to the highest stellar density regions. In this section, we use the extinction law derived in the main body of the paper to map the $K_s$ extinction using two methods: (i) the red clump method and (ii) the Rayleigh Jeans colour excess method.

\subsection{Red clump method}\label{appendix::rc_method}
A standard method for measuring the mean extinction for a region of the sky is to find the $(J-K_s)$ colour of the red clump and attribute the difference with respect to a standard value $(J-K_s)_0$ as due entirely to extinction \citep{Gonzalez2011,Gonzalez2012,Surot2020}. For a population of stars with $(J-K_s)>0.5$, within a colour-magnitude cut encompassing the red clump $11.5<K_s-0.482(J-K_s)<14$ and with magnitude uncertainties $<0.2\,\mathrm{mag}$, we find the colour of the red clump as the location of the maximum density from a multi-Gaussian fit (using the Bayesian information criterion to choose the number of components) and measure its width from the full-width at half maximum. We avoid misidentifying the overdensity due to young turn-off stars by ensuring the peak is always located at $<20\percent$ the colour range for $(J-K_s)$ at $|b|<2\,\mathrm{deg}$. The choice of $(J-K_s)_0$ requires some care for precision work. \cite{Gonzalez2012} use $(J-K_s)_0=0.68$ based on measurements of extinction in other bands in Baade's window and using an assumed extinction law, whilst \cite{Simion2017} use $(J-K_s)_0=0.62$. We opt to `zero-point' the $E(J-K_s)_\mathrm{RC}$ measurements using the sample of RR Lyrae from \cite{Molnar2021}. We select high-confidence RR Lyrae (classification probability $>0.9$) within $|\ell|<1.5\,\mathrm{deg}$ and $|b|<1.5\,\mathrm{deg}$ and with $E(J-K_s)_\mathrm{RC}<3$ to avoid biases at high extinction discussed below (assuming initially $(J-K_s)_0=0.62$ but this does not matter). We compute the sample's unextincted $(J-K_s)$ using the period-luminosity-metallicity relations from \citet[][also derived on the VISTA system]{Cusano2021} assuming [Fe/H]$=-0.94\,\mathrm{dex}$ \citep[][although the gradient of $(J-K_s)$ with metallicity is $0.007\,\mathrm{mag}\,\mathrm{dex}^{-1}$ so varying the metallicity within reasonable values has a very weak effect]{Dekany2021}. The peak distance of this sample is $ 8.22\,\mathrm{kpc}$ (assuming $(J-K_s)_0=0.62$ and $A_{Ks}/E(J-K_s)=0.44$ as found in the main body of the paper) giving confidence that these RR Lyrae are behind a similar level of extinction to the red clump stars. We find $(J-K_s)_\mathrm{RC}-E(J-K_s)_\mathrm{RRL}\equiv(J-K_s)_0=(0.647\pm0.004)$. This is redder than the solar neighbourhood $(J-K_s)_0=(0.588\pm0.006)$ from \citet[][transformed to the VISTA system using the equations from \citealt{GonzalezFernandez2018}]{ChanBovy2020} due to the correlation between effective temperature and metallicity for red clump stars.

For regions of high extinction, the $J$ photometry of red clump stars at the Galactic centre becomes incomplete and so extinction estimates via this method are biased low. In these regions $(H-K_s)$ is a better extinction estimator although in lower extinction regions it lacks the dynamic range of $(J-K_s)$. We employ the same procedure for $(H-K_s)$ with a selection $(H-K_s)>0$ and $11.5<K_s-1.13(H-K_s)<14$, avoiding the turn-off by ignoring peaks $<10\percent$ for $(H-K_s)$ for $|b|<1.5\,\mathrm{deg}$. We cannot zero-point the $(H-K_s)$ excesses using the RR Lyrae sample as no $H$-band period-luminosity relation is available from \cite{Cusano2021}. We instead reference with respect to $E(J-K_s)_\mathrm{RC}$ at low extinction assuming $E(J-K_s)/E(H-K_s)=2.879$ from Table~\ref{tab:colour_ratio_results}. Restricting to stars with $E(J-K_s)_\mathrm{RC}<3$ to avoid the underestimates at high extinction, we find $(H-K_s)_0=(0.142\pm0.001)$. This agrees very well with the median $(H-K_s)=(0.130\pm0.001)$ of low extinction ($A_{Ks}<0.01$) APOGEE DR17 \citep{APOGEE1,APOGEE2} stars with $2.3<\log g<2.5$ and $0.63<(J-K_s)<0.67$ using the transformations from 2MASS to VISTA from \cite{GonzalezFernandez2018}.

Another consideration is resolution: extinction often varies on very small scales but we must average over sufficiently large patches of sky to yield enough red clump stars. However, because of extinction, the number density of stars varies on small scales making a fixed resolution inappropriate. \cite{WeggGerhard2013} use the \cite{Gonzalez2011} method employing an adaptive resolution. Here we follow a similar procedure using the properties of the nested Healpix scheme \citep{Fernique2014}. We first bin stars using a Healpix NSIDE of $256$ (angular resolution of $13.7\,\mathrm{arcmin}$) and subdivide each bin individually until any more subdivision would produce fewer than $100$ stars per bin. This produces resolutions down to $0.3\,\mathrm{arcmin}$. The advantage of using the nested Healpix scheme is that each index can be simply converted to an index at the highest resolution of the map. A query for the extinction at a given location is then a fast binary search for where the index of the required Healpix sits within the list of sorted pixels. The extinction is that of the entry below the insertion point. The resolution of our maps is inferior to the recent work from \cite{Surot2020} who use a similar method but require only $20$ stars to produce an extinction estimate. The resulting maps are consistent with ours but have a larger scatter from using fewer stars.

\begin{figure*}
\includegraphics[width=.99\linewidth]{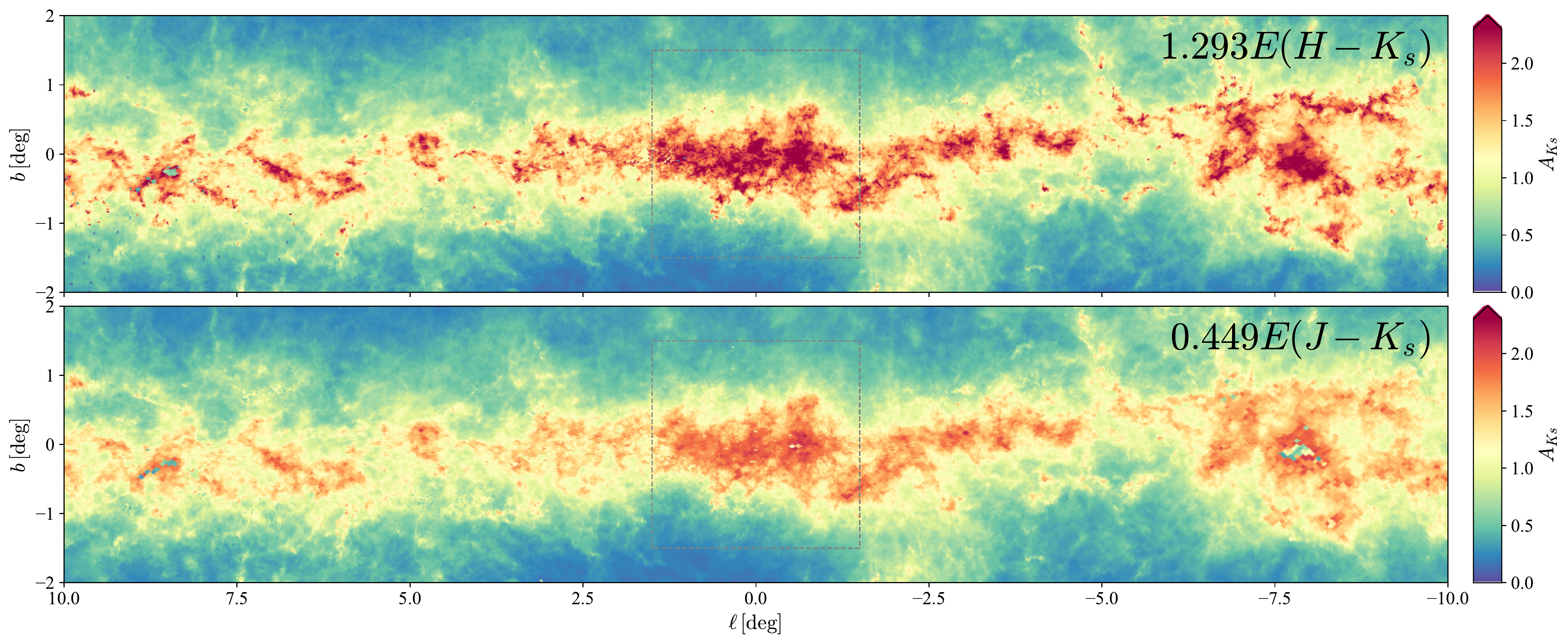}
\includegraphics[width=.99\linewidth]{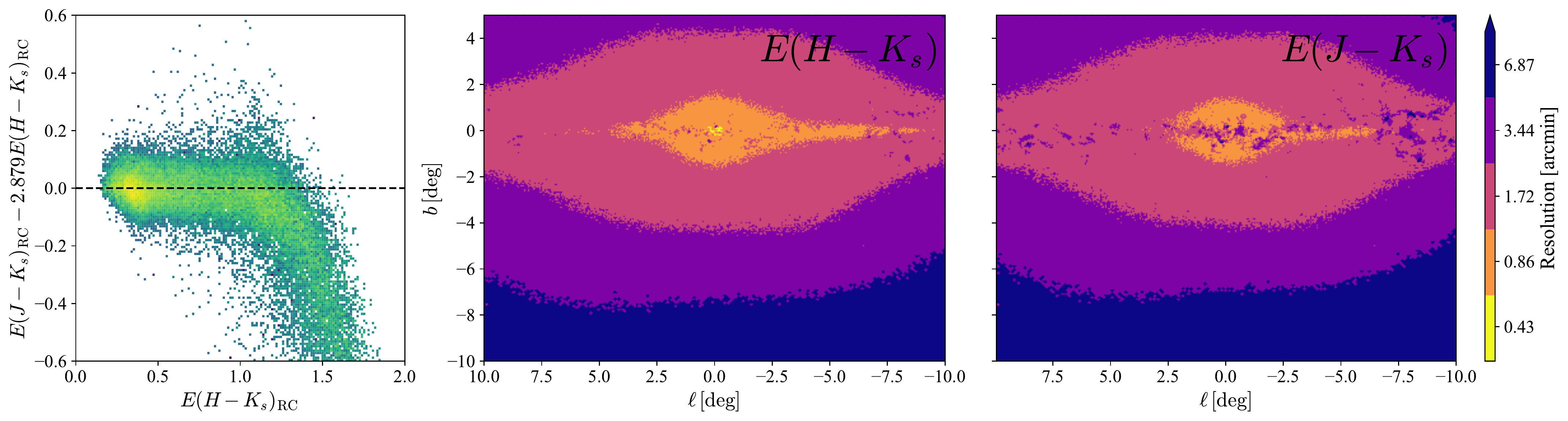}
\caption{Adaptive resolution extinction maps. The top two panels show the $A_{K_s}$ extinction computed from the $(H-K_s)$ and $(J-K_s)$ colours of the red clump stars (the grey box gives the region studied in this paper). The bottom left panel shows the difference between the map pixels assuming $E(J-K_s)/E(H-K_s)=2.879$ as per Table~\ref{tab:colour_ratio_results}. The high extinction regions have underestimated $E(J-K_s)$ as the Galactic centre/bulge red clump stars are lost in VIRAC2 $J$ in these regions. The right two panels in the bottom row shows the adaptively chosen resolution of the maps such that $>100$ red clump stars are in each pixel.}
\label{fig:ext1to1}
\end{figure*}

In Fig.~\ref{fig:ext1to1} we display the $K_s$ extinction maps within the bulge midplane $|b|<2\,\mathrm{deg}$ derived from the red clump $(H-K_s)$ and $(J-K_s)$ colours along with the resolution maps across the entire bulge. The extinction coefficients are taken from Table~\ref{tab:ext_results}. We observe excellent correlation between the two maps in lower extinction regions. It is clear the $(J-K_s)$ map underestimates the extinction in regions of high extinction due to the red clump stars being too faint in $J$ in these fields. In many cases, foreground red giants with lower extinction are identified causing slight underestimates, but in extreme cases (e.g. $(\ell,b)=(-8,0)\,\mathrm{deg}$) the algorithm has possibly identified young main sequence stars as red clump stars leading to severe underestimates. The resolution maps largely trace the stellar density (note the bar-bulge asymmmetry). In low-latitude, high-extinction regions, the $J$ resolution is decreased.

\subsection{Rayleigh Jeans Colour Excess method}\label{appendix::rjce_method}
As with the $J$ photometry, $H$ of the bulge red clump becomes incomplete in some high extinction regions, producing biased results. In these regions we can employ an alternative method that uses all red giant stars so is less reliant on observing the red clump. \cite{Majewski2011} introduced the `Rayleigh Jeans Colour Excess' method by showing how red giants have a near constant intrinsic $(H-[4.5])$ colour such that the extinction can be estimated as $A_{Ks}=0.918(H-[4.5]-0.08)$ based on the extinction law from \cite{Indebetouw2005}. \cite{Soto2019} used this technique with VVV and GLIMPSE data in the southern disc. Recently, \cite{StelterEikenberry2021} have used the RJCE method on integrated photometry of the Galactic centre region finding an infra-red slope of $\alpha=(2.03\pm0.06)$.
We use the sample of VIRAC2 sources cross-matched to the GLIMPSE catalogue \citep{GLIMPSE} and the Spitzer-IRAC GALCEN point source catalogue of \cite{Ramirez2008} using a $0.4\,\mathrm{arcsec}$ cross-match radius, as detailed in Section~\ref{sec::data}. This uses stars with $[4.5]$ magnitude uncertainties $<0.2$, $H$ uncertainties $<0.06$, no neighbour within $1\,\mathrm{arcsec}$, no potential AGB or YSO stars (using $([5.8]-[8.0])$) and the cuts on extinction-corrected $(H-K_s)$ and $(J-K_s)$ (when available) to remove potential foreground contaminants. We have calibrated the relationship between $E(H-K_s)$ and $(H-[4.5])$ using the method in Section~\ref{section::extinction}. We fix $E(H-[4.5])/E(H-K_s)$ using the coefficient from Table~\ref{tab:colour_ratio_results} and fit for the intercept of $(H-[4.5])$ vs. $E(H-K_s)_\mathrm{RC}$ accounting for non-linearities and the giant branch slope. We find $E(H-[4.5])=0.967(H-[4.5]+0.0482(K_s-12.514)-0.210)$. Note the non-unity leading coefficient is due to the slight gradient of the giant branch in $(H-[4.5])$. Combining with the $A_{Ks}/E(H-[4.5])$ found in the main body of the paper we find
\begin{equation}
    A_{Ks}=0.677(H-[4.5]-0.188),
\end{equation}
for $K_s=13$ stars. The zero-point of $(H-[4.5])_0=0.21$ is redder than that used by \cite{Majewski2011} based on the \cite{Girardi2002} isochrones. This could be linked to zeropoint issues in the photometry or population effects, particularly as the red clump stars in this region also appear intrinsically redder in $(J-K_s)$ and $(H-K_s)$ than in the solar neighbourhood. For a set of Healpix with NSIDE of $8192$ ($26\,\mathrm{arcsec}$ resolution), we find the $100$ nearest neighbours to the centre of the pixel and measure the median $E(H-[4.5])$ colour and its spread. 

\begin{figure*}
    \centering
    \includegraphics[width=\textwidth]{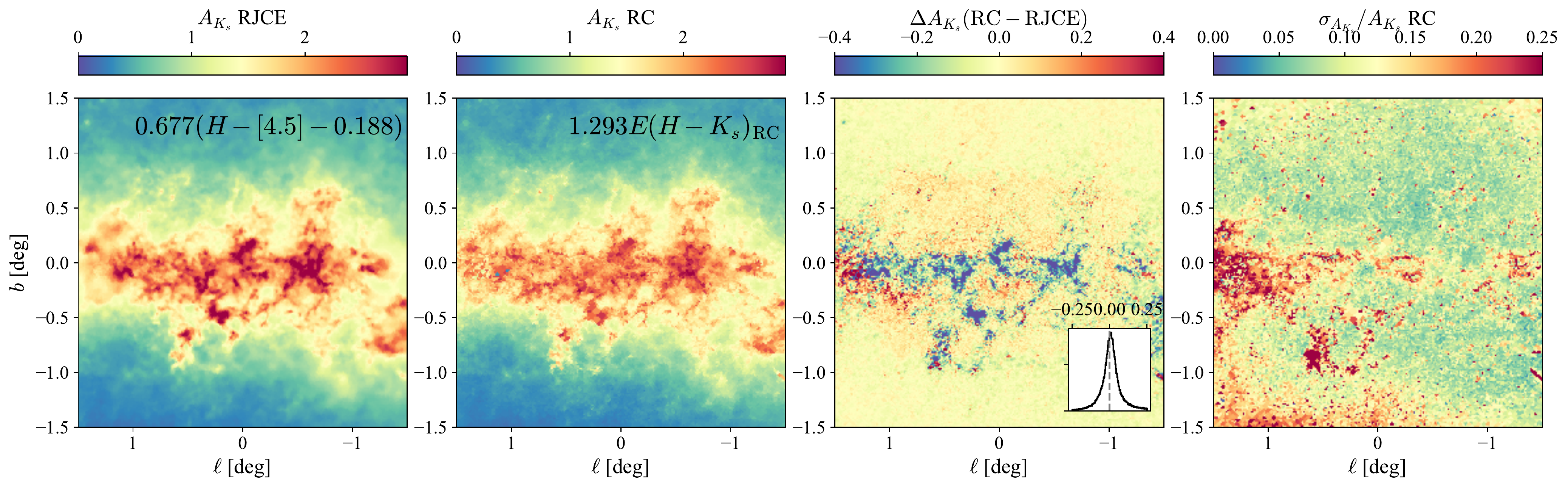}
    \caption{Comparison of extinction maps computed using the RJCE method (first panel) and the red clump method (second panel). The difference is shown in the third panel. Note the good correspondence except within the very high extinction regions where the bulge red clump is not visible in VIRAC2 $H$ so extinction is underestimated by the red clump method. Note also the general overestimate in the central $-1\lesssim\ell\lesssim1$ and $-0.7\lesssim b\lesssim0.7$ potentially due to zero-point differences between the two Spitzer surveys, \protect\cite{Ramirez2008} and \protect\cite{GLIMPSE}. The right panel shows the relative uncertainty in the red clump extinction estimates. }
    \label{fig::rjce_comparison}
\end{figure*}

In Fig.~\ref{fig::rjce_comparison} we display a comparison of the maps derived using the RJCE method with those from the red clump method. We see from the difference map that on the average the maps match well, but in the high extinction regions of the mid-plane the red clump method significantly underestimates the extinction. For the high extinction regions the red clump stars in the centre of the Galaxy are too faint in $H$ for VIRAC2 so the average $E(H-K_s)_\mathrm{RC}$ extinction measurement is to foreground red clump stars. There is a general discrepancy (faint red/orange) in the central $-1\lesssim\ell\lesssim1$ and $-0.7\lesssim b\lesssim0.7$ corresponding to the footprint of the \cite{Ramirez2008} survey with the RJCE method underestimating the extinction relative to the RC method. \cite{Ramirez2008} report their $[4.5]$ measurements being $0.06\,\mathrm{mag}$ larger in the mean than the GLIMPSE measurements in the overlapping regions consistent with this bias. We also note the region around $(\ell,b)=(1.25,0)^\circ$ where the RJCE method underestimates the extinction. It appears that here there is a significant contribution from young stars. This suggests there is important variation of the extinction along the line-of-sight for this region and highlights the limitations of our method. We can see that clearly in the map of the relative spread in extinction from $(H-K_s)$ red clump method (the spread from the RJCE method displays a similar feature) where this region is one of the most significant, along with the feature at $(\ell,b)=(0.5,-0.8)^\circ$.

\section{Completeness of the VIRAC2 catalogue}\label{section::completeness}

\begin{figure}
    \centering
    \includegraphics[width=\columnwidth]{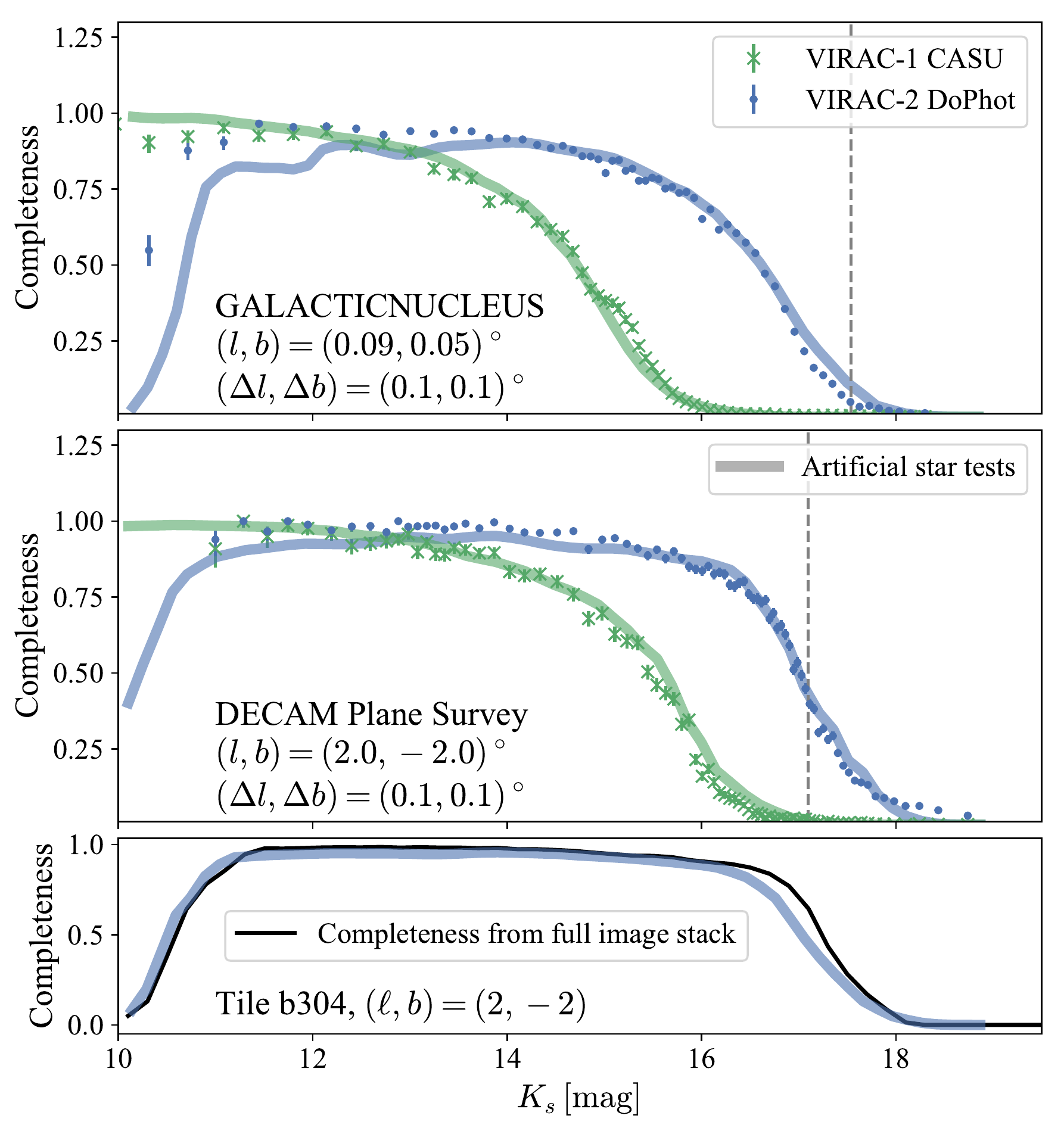}
    \caption{VIRAC completeness tests: the top two panels show the completeness for two $0.1\,\mathrm{deg}$ by $0.1\,\mathrm{deg}$ fields centred on $(\ell,b)=(0.19,0.18)\,\mathrm{deg}$ (top) and $(\ell,b)=(4,2)\,\mathrm{deg}$ (bottom). Solid lines show the expected completeness from artificial star tests and the points with errorbars (blue dots for VIRAC2 which uses DoPhot and green crosses for VIRAC v1 which uses CASU's aperture photometry program \emph{imcore}) show the completeness measured with respect to the GALACTICNUCLEUS (top) and DECAPS catalogues (bottom). The dashed lines show the magnitudes at which the magnitude distributions of GALACTICNUCLEUS and DECAPS turn over. The bottom panel shows a check of our simple completeness procedure using a representative set of images (thick blue) compared to the results of analysing a full image stack (thin black).}
    \label{fig:completeness_test}
\end{figure}

In this work we have measured the density profile of red clump stars. For this, we require knowledge of the incompleteness of the source catalogue i.e. the probability of a source with a given magnitude and on-sky location being detected under given observing conditions. Incompleteness arises from seeing and sky brightness limitations which alter the limiting magnitude depth of a survey and can result in blending in crowded regions. Traditionally, incompleteness has been assessed through two methods. In some cases, a deeper, more complete catalogue is available to which we can exactly compare which sources in our catalogue of interest were detected. However, for the VVV survey, no such deeper catalogue is available across the entirety of the survey footprint and so we must instead estimate the source recovery rate another way. One method for assessing this is through testing the recovery of artificial stars injected into the reduction pipeline.  VIRAC v1 used aperture photometry for which \cite{Saito2012} performed these artificial star tests. Here we will follow this method by checking the recovery of artificial stars using point-spread-function photometry utilised in VIRAC2 and we will use the method of comparison to deeper catalogues as a cross-check in survey regions overlapping deeper surveys.

In a stacked image (a combination of two dithered images) from a single detector, we inject artificial stars arranged on a regular hexagonal grid spaced by $30$ pixels (typical seeing is $\sim 2.2$ pixels). This results in $\sim 5700$ stars being injected. The stars are assigned magnitudes randomly sampled between $K_s=10\,\mathrm{mag}$ and $K_s=20\,\mathrm{mag}$, and follow a Gaussian point-spread-function with full-width at half-maximum equal to the seeing stored in the header for each image. The fake image is processed using $\textsc{DoPhot}$ using the same configuration parameters as adopted by Smith et al. (in prep.). An artificial star is `recovered' if there is an output source approximately within $1$ pixel ($0.339\,\arcsec$ assessed using a k-d tree) and $1$ mag of the input, and the source also lies within a region of the associated CASU confidence map \citep{Emerson2004} with value $>25$. We store the fraction of sources recovered in bins of $0.2\,\mathrm{mag}$ and apply a gaussian smoothing filter of one bin width. 
This procedure is repeated for all $16$ detectors, although we ignore the results from detector $4$ which is known to have a number of bad rows (see \href{http://casu.ast.cam.ac.uk/surveys-projects/vista/technical/known-issues#section-5}{the CASU webpages}). Note that our procedure ignores completeness variations within a single detector image, which could arise due to variations in the detector properties or variations in crowding. This is a particularly noticeable issue near the nuclear star cluster where there is large density variation on a sub-detector level. Under the assumption of uniform density, our completeness estimates will, in general, underestimate the completeness correction slightly as more stars lie in higher crowding regions which have lower completeness. In this way, the star-weighted completeness over a detector image is lower than that estimated from our uniform distribution of artificial stars.

The VIRAC2 catalogue is a multi-epoch astrometric catalogue so the probability of a source entering the catalogue depends on the probability of detection in every image. Typically \textsc{DoPhot} reports a number of false detections, which can arise, for instance, in the wings of bright sources. Therefore, reliable sources require detection in multiple images, or other quality cuts. However, performing the artificial star tests on every image is a costly procedure, and many images will have very similar completeness properties. We therefore opt to only perform the artificial star tests on representative sets of images. Two typical quality cuts employed on the VIRAC2 dataset are requiring a five-parameter astrometric solution (at least $10$ epochs -- typically fields have $>100$ epochs) and/or that the source is detected in more than $20\percent$ of the observations. For each unique bulge tile and pointing combination ($6$ pointings per tile), we perform our procedure on the images at the $2$nd, $20$th and $80$th percentiles of seeing. We choose the $2$nd as this represents the `best' completeness, $20$th as it corresponds to at least $20\percent$ detection rate at the faint end and $80$th to correspond to at least $20\percent$ detection rate at the faint end. Each set includes $196\times6\times16=18,816$ individual detector images so we process a total of $\sim55,000$ images. We combine the results of the $20$th and $80$th percentile results taking the maximum completeness at each $K_s$, and use these values as our standard completeness values.

The resulting catalogue of completeness ratios as a function of $(\ell,b,K_s)$ is interpolated onto a regular grid using inverse distance weighting of $5$ nearest neighbours. This regular grid is then interpolated using cubic splines for any required point. We perform a cross-check of our procedure by adding a set of artificial stars into all images in a single tile: b304 centred around $(\ell,b)\approx(-2,-2)\,\mathrm{deg}$. We first create a hexagonal grid of sources in equatorial coordinates spaced by $30$ VIRCAM pixels ($10.2\,\mathrm{arcsec}$) over the area covered by the tile (approximately $260,000$ stars) and assign random magnitudes as before. For each detector image we insert the stars at the appropriate pixel locations, process with DoPhot and check recovery as before. We record the total number of observations and detections for each artificial source. In the lower panel of Fig.~\ref{fig:completeness_test} we show the ratio of the number of artificial sources with more than $20\percent$ detections to the total number and compare to the expected completeness from the previous approach. We note the satisfactory agreement. From $K_s\approx12$, our previous approach slightly underestimates the true completeness and this is most noticeable at the faint end. It appears this is due to the incorrect assumption that the completeness in the regions covered by overlapping images with similar seeing are equal in the two images. Particularly for fainter magnitudes, a similar fraction of sources is recovered in each image, but in the overlapping sections it is a slightly different set of sources. This means more sources in total are detected, slightly increasing the completeness fraction. Possibly this is related to varying performance across each detector. At the bright end, the well-matched turnover validates our procedure for combining the results of the $20$th and $80$th percentiles.

\begin{figure}
    \centering
    \includegraphics[width=\columnwidth]{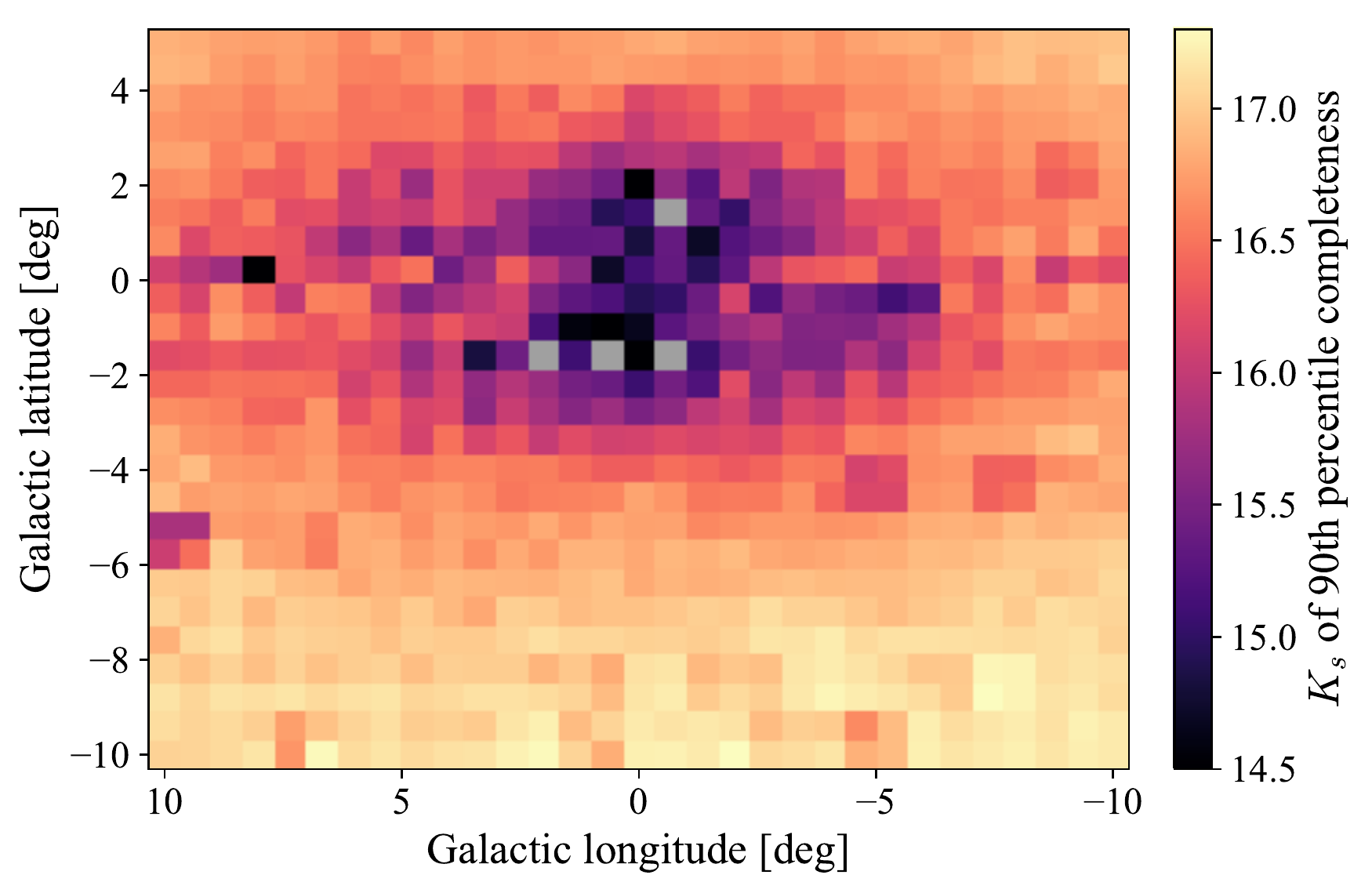}
    \caption{$K_s$ magnitude at which the bulge region of VIRAC2 is $90\percent$ complete. The grey pixels are regions where $90\percent$ completeness is not reached for any magnitude.}
    \label{fig:completeness90}
\end{figure}

In Fig.~\ref{fig:completeness90} we show the on-sky distribution of the $K_s$ magnitude at which the bulge region of VIRAC2 is $90\percent$ complete. We see at high latitude the completeness is above $90\percent$ down to magnitudes fainter than $K_s=17$. For higher density regions near the midplane and inner bulge (the subject of this paper) the completeness is above $90\percent$ only below $K_s\approx15.5$.

\subsection{Comparison with external catalogues}
We validate our procedure by inspecting the expected completeness ratio against that measured by comparison to the GALACTICNUCLEUS survey \citep{GALACTICNUCLEUS} and the DECAM Plane Survey \citep{Schlafly2018}. The GALACTICNUCLEUS survey is a $JHK_s$ imaging survey using HAWK-I on the VLT reaching a $5\sigma$ depth of $K_s\approx21$. The survey has $49$ pointings of $7.5'$ by $7.5'$ primarily located within the nuclear stellar disc, but several pointings lie slightly out of the midplane. \cite{GALACTICNUCLEUS} estimates the catalogues are $\sim80\percent$ complete at $K_s=16\,\mathrm{mag}$. The difference between $K_s$ measurements for VIRAC2 vs. GALACTICNUCLEUS is $0.06\,\mathrm{mag}$ for the field inspected below. DECAM Plane Survey (DECAPS) is a $grizY$ imaging survey using the Dark Energy Camera at Cerro Tololo covering $-120<\ell/\,\mathrm{deg}<5$ and $|b|\lesssim4\,\mathrm{deg}$. The survey has a $6\sigma$ depth in $Y$ of $21$. We compute an approximate $K_s$ magnitude from $izY$ using $K_s\approx i-3.34(i-z)-1.98(z-y)-0.69$ (this only works well for low extinction regions -- the region we choose has $A_{K_s}\approx0.2\,\mathrm{mag}$).

Fig.~\ref{fig:completeness_test} shows the completeness of two fields with respect to the GALACTICNUCLEUS and DECAPS catalogues alongside the expectation from the artificial star tests. We display the ratio of sources that have a cross-match in VIRAC2 as a function of $K_s$ (using the approximate $K_s$ for DECAPS). The two fields are $0.1\,\mathrm{deg}$ by $0.1\,\mathrm{deg}$ centred on $(\ell,b)=(0.19,0.18)\,\mathrm{deg}$ and $(\ell,b)=(4,2)\,\mathrm{deg}$. In these fields, there are respectively $\sim670,000$ and $\sim 2.3$ million sources per square deg with $K_s<16$ in VIRAC2. We note the very good correspondence between expected and measured completeness. We naturally expect small differences as the parent catalogues (GALACTICNUCLEUS and DECAPS) are not perfectly complete. Indeed $2$ and $5\percent$ of sources in the VVV catalogue are not cross-matched within $0.4\,\mathrm{arcsec}$ to any source in the GALACTICNUCLEUS and DECAPS catalogue respectively. Additionally, we show the magnitudes at which the magnitude distributions of GALACTICNUCLEUS and DECAPS begin turning over ($\sim17.5$ and $\sim17$ respectively). This magnitude is a good indicator of the point at which a catalogue starts becoming incomplete. For both comparison fields, our completeness calculations match the recovery ratio very well. We also display the equivalent results for the VIRAC v1 catalogue which used the CASU aperture photometry software \emph{imcore}. In the DECAPS field there is $\sim1$ magnitude increase in the $50\percent$ completeness level of VIRAC2 compared to v1 which increases to $\sim2$ magnitudes for the highly crowded GALACTICNUCLEUS field.

\bsp	
\label{lastpage}
\end{document}